\documentclass[prd,aps,showpacs,preprintnumbers,amssymb]{revtex4}

\usepackage{epsf}

\def\e3p{$\eta \rightarrow 3 \pi$}

\begin{document}

\title{%
\hfill{\normalsize\vbox{%
\hbox{\rm SU-4252-845}
}}\\
{Two chiral nonet model with massless quarks}}

\author{Amir H. Fariborz $^{\it \bf a}$~\footnote[3]{Email:
fariboa@sunyit.edu}}

\author{Renata Jora $^{\it \bf b}$~\footnote[2]{Email:
cjora@physics.syr.edu}}

\author{Joseph Schechter $^{\it \bf c}$~\footnote[4]{Email:
schechte@physics.syr.edu}}

\affiliation{$^ {\bf \it a}$ Department of Mathematics/Science,
State University of New York Institute of Technology, Utica,
NY 13504-3050, USA.}

\affiliation{$^ {\bf \it b,c}$ Department of Physics,
Syracuse University, Syracuse, NY 13244-1130, USA,}

\date{\today}

\begin{abstract}

   We  present a detailed
study of a linear sigma model containing one chiral nonet
transforming under U(1)$_A$ as a
quark-antiquark composite and
another chiral nonet transforming as a
diquark-anti diquark composite
(or, equivalently from a symmetry point of view, as
a two meson
molecule).
The model provides an intuitive explanation of a
current puzzle in low energy QCD: Recent work has
suggested the existence of a lighter than 1 GeV
nonet of scalar mesons which  behave like four quark
composites. On the other hand, the validity of a
spontaneously broken chiral symmetric description
would suggest that these states be
chiral partners of the light
pseudoscalar mesons, which are two quark composites.
The model solves the problem by starting with the
two chiral nonets mentioned and allowing them to mix
with each other. The input of physical masses
in the SU(3) invariant limit
for two scalar octets and an ``excited" pion octet
results in a mixing pattern wherein the light scalars
have a large four quark content while the light
pseudoscalars have a large
two quark content. One light isosinglet
scalar is exceptionally light. In addition, the
pion pion scattering is also studied and the
current algebra theorem is
verified for massless pions which  contain some
four quark admixture.
\end{abstract}

\pacs{13.75.Lb, 11.15.Pg, 11.80.Et, 12.39.Fe}

\maketitle

\section{Introduction}
    The topic of anomalously light scalar mesons in QCD has become a
subject of increasing interest in the last fifteen years or so
\cite{ropp}-\cite{liu}. Of course light scalars, especially the
light isoscalar called first sigma and now $f_0(600)$, have been
discussed for
at least three times as long, although without general agreement
on their actual existence.
The difficulty
people had previously in accepting the light scalars was largely due to
the great success of the simple quark model, in which the lightest
scalars are expected to be p-wave quark- antiquark composite states and
hence to be in the
1 to 1.5 GeV range, like the other p-wave states.
It seems that physicists now believe more in their
existence because there have been an increasing number of investigations,
using a variety of techniques and models, which suggest that they do
exist. Common features in many of these approaches have been the use of
unitarity (which no one denies) and some input at low energy from
chiral dynamics (which is also considered reasonable).

    Of course, the strongly interacting gauge theory QCD
has not been ``solved" and any possible new features in the low
energy region where the effective coupling constant is especially
strong  raise the hope of improving one's understanding of this
basic theory. Perhaps the most fascinating possibility is that
the very light scalars contain two quarks and two antiquarks.
Variants based on a diquark- anti diquark picture \cite{j}
or a meson-meson
``molecule" picture \cite{iw} have been discussed.

    A lot of attention has been given to the question of a
possible nonet grouping for the light (less than 1 GeV) scalars.
The candidates are the already mentioned  $f_0$(600),
the Kappa(800-900)
[not conclusively established
according to \cite{ropp}], the established $a_0$(980) and the
established $f_0$(980). It has been pointed out
[See for examples \cite{j}, \cite{BFSS2} and \cite{MPPR04}] that a
characteristic
signature of a four quark content would be an inverted mass
ordering, with an almost degenerate $I=0$, $I=1$ pair being the heaviest
rather than the lightest states when the light quark masses are
`` turned on." This seems to be the case.

     Associating the four quark states with the lightest scalars
naturally raises the question of where are the p-wave quark, antiquark
scalars. The candidates for the non-zero isospin states are the
established $a_0$(1450) and the established $K_0^*$(1430). For the
$I=0$ states the established candidates are the $f_0$(1370),
$f_0$(1500) and the $f_0$(1710), one of which may be a glueball.
There is a slight puzzle since the non strange $a_0$(1450)
with a listed mass of 1474 MeV is heavier than the strange
$K_0^*$(1430) with a listed mass of 1414 Mev. In
addition some branching
ratios are not well predicted by SU(3) invariance.
A possible way to
overcome this problem [\cite{BFS3}, \cite{mixing}]
is to allow mixing between the lighter 4
quark and heavier 2 quark scalar nonets. This feature
is incorporated
as a basic part of the present paper.

   While it is rather difficult to treat low energy QCD dynamically,
great success has been obtained at low energies using the underlying
chiral symmetry of QCD. We also incorporate this as an aid in getting
more information about the system. This feature will be implemented
by using linear rather than the more usual nonlinear representations
for the pseudoscalar and scalar fields. If both four quark and two quark
scalars are present, this means that four quark pseudoscalars should also
appear in the model. Experimental candidates for the non-zero isospin,
higher mass pseudoscalars are the $\pi(1300)$ and the two not yet
conclusively established strange states $K(1460)$ and $K(1830)$.
The candidates for the higher mass isoscalar pseudoscalars are the
$\eta(1295)$, $\eta(1405)$, $\eta(1475)$ and the not conclusively
established $\eta(1760)$. It is possible that one or more of these
experimental candidates also contain glueball and radial excitation
admixtures. At first glance it might seem puzzling that the
picture seems to be:
light mass two quark and heavy mass four quark states for the
pseudoscalars
at the same time as light mass four quark and heavy mass two
quark states for their ``chiral partners," the scalars. We shall make
no initial assumption on this matter but let the experimental
particle spectrum together with the mixing inherent in the model
tell us the answer.

     At a technical level it is amusing to note that the
U(1)$_{\rm A}$ transformation properties distinguish the two quark from
the four quark
fields. Since it is known that the U(1)$_{\rm A}$ symmetry is badly broken
in QCD, this means that we have to model the breaking in some detail.
For this purpose we will use an extra term in addition to the usual one.
We adopt a  counting scheme for selecting the most important
terms, out of the many possible ones. We assign a number $N$
equal to the number of underlying quark plus antiquark lines
associated with each effective term. Then
it seems reasonable to pick up the terms with smallest $N$ values.
On this basis the extra term for saturating the  U(1)$_{\rm A}$ anomaly
has the same justification as the conventional one.

     Clearly, with so many scalar and pseudoscalar fields present, the model
is fairly complicated to analyze. At the same time it is widely believed
that massless (ie zero mass light quarks) QCD is an excellent
qualitative approximation. Except for the pseudoscalar Nambu Goldstone
bosons of the theory, the masses of the physical particles made from light
quarks are largely due to the spontaneous breakdown of chiral symmetry.
We will employ this limit of the theory in the present paper and note that 
it
very much simplifies the analysis. Especially, the characteristic
mixing matrix pattern
of the two quark and four quark states becomes very clear.
The puzzle of opposite two quark vs four quark structures
of the scalars and pseudoscalars seems to be neatly solved by the
mixing mechanism.

    Even though the nonlinear chiral model is more convenient for
systematically studying the loop corrections at very low energies, the
linear sigma models have a long history of elucidating key features of
the strong interactions. Roughly speaking the use of the  nonlinear model
amounts to integrating out the scalars (although it is technically
somewhat more general than that). Certainly for learning about the scalars
themselves it is rather convenient to have them present in the Lagrangian
to begin with. The most famous example of the linear model is of course the
Higgs potential of the standard model. One of the classical triumphs of the
nonlinear model is the derivation of the ``current algebra" formula for
low energy pion scattering. This can be obtained, though in a more
complicated
way, also in the linear model. We verify this in detail in the present 
paper.
One might wonder, since the pion in the present model has a small
(but non negligible) four quark content, whether the current algebra result
strictly does hold in the present model. Our result shows that it does hold 
for
the zero pion mass case we are considering here.

    The two chiral nonet model was introduced in \cite{BFMNS01} as a
convenient way to study the possibility of mixing
between quark- antiquark ($q{\bar q}$) spin zero mesons and
two quark- two antiquark ($qq{\bar q}{\bar q}$) spin zero mesons.
Altogether there are two pseudoscalar and two scalar nonets
contained in the model. It was found that, in the zero quark mass limit
with just a few explicit chiral invariant terms contained, there was a
possibility of a situation in which the lightest pseudoscalars could have
zero mass (i.e. be Nambu-Goldstone bosons) and be primarily
$q{\bar q}$ type while
the next heaviest mesons could be scalars, primarily of
$qq{\bar q}{\bar q}$ type.
Furthermore, the next heaviest mesons could be pseudoscalars of
mainly four quark type while the heaviest could be scalars, mainly of two
quark type.
A treatment \cite{NR04} of the model with similar chiral
invariant
terms and several different quark mass terms also found that light
scalars with relatively large admixtures of $qq{\bar q}{\bar q}$
type states are favored.
Actually, the model can be rather complicated
since there are twenty one renormalizable
chiral invariant terms which can be
made as well as a similar number of renormalizable
quark mass type terms which transform
as the $(3,3^*)+(3^*,3)$ representation of chiral SU(3)$_{\rm L} \times$
SU(3)$_{\rm R}$.
In \cite{FJS05}, the present authors
studied the more general version of the model in which all
possible chiral invariant, even non renormalizable,
terms were included together with the single usual realization of the
quark mass term. The same overall picture was found.
However, because the
method
relied on the symmetry properties of the Lagrangian, only the
properties of the
pseudoscalar states and the strange scalar states could be studied.
In the present paper, we shall initiate a much more systematic
investigation. We first study precisely how the general results get
constrained when a specific choice of invariant interaction
terms is made. We introduce a scheme for ordering all
the non-derivative terms of the
Lagrangian according to their likely importance. This
enables us to select a limited number of leading order terms
in a meaningful way as well as to provide the framework for
possible higher order extensions.
At leading order, and with the extra simplification
of zero quark masses, all our results were determined analytically,
without any need for a numerical fitting procedure.
There are essentially only four main input parameters
and only one of them
has a non negligible experimental error. We do our
calculations for all
allowable values of this parameter $(m[\pi(1300)])$
and also take into account the small experimental error
on another of the
three parameters.
The results obtained dramatically predict the existence
of a very low mass scalar isosinglet state. Especially,
the puzzle concerning the coexistence of lighter (mainly) two quark
pseudoscalars with lighter (largely) four quark scalars is
clearly seen to be solved.

     A brief review of the model and the relevant notation is presented
in Sec. II. Section III shows the great simplifications obtained by
going to massless QCD and also gives our notations appropriate to the
flavor SU(3) invariant situation in this limit. General results, valid
for any choice of terms in the invariant potential, are also presented
in this section. Section IV gives a systematic procedure for
deciding  which terms are most important in the model. It mainly contains
the worked out model using the leading terms in this scheme. A numerical
analysis is presented and the masses of the two SU(3) singlet scalar states
of the model are predicted. The two and four quark contents for
each state of the model are displayed.
Sections V, VI and VII are devoted to proving,
for any choice of invariant potential, the current algebra theorem for
the scattering of massless pions. Discussion and conclusions are given in
Sec. VIII.

\section{Brief review of model}

    The fields of our ``toy" model consist of a 3 $\times$ 3 matrix chiral
nonet
field, $M$ which represents $q{\bar q}$ type states as well as a
3 $\times$ 3 matrix chiral nonet
field, $M^\prime$ which represents $qq{\bar q}{\bar q}$ type states.
      They have the
decompositions into scalar and pseudoscalar pieces:
\begin{eqnarray}
M &=& S +i\phi, \nonumber  \\
M^\prime &=& S^\prime +i\phi^\prime.
\label{sandphi}
\end{eqnarray}
They behave under ``left handed" and ``right handed" unitary
unimodular (ie SU(3)$_{\rm L}\times$ SU(3)$_{\rm R}$) transformations  as
\begin{eqnarray}
M \rightarrow U_L M U_R^\dagger, \nonumber  \\
M^\prime \rightarrow U_L M^\prime U_R^\dagger.
\label{Mchiral}
\end{eqnarray}
However, under
the U(1)$_{\rm A}$ transformation which acts at the quark level as
$q_{aL}
\rightarrow e^{i\nu} q_{aL}$, $q_{aR} \rightarrow e^{-i\nu} q_{aR}$, the
two fields behave differently:
\begin{eqnarray}
M \rightarrow e^{2i\nu} M, \nonumber  \\
M^\prime \rightarrow e^{-4i\nu} M^\prime.
\label{MU1A}
\end{eqnarray}
We will be interested in the situation where non-zero vacuum values
of the diagonal components of $S$ and $S'$ may exist. These will be
denoted by,
\begin{equation}
\left\langle S_a^b \right\rangle = \alpha_a \delta_a^b,
\quad \quad \left\langle S_a^{\prime b} \right\rangle =
\beta_a \delta_a^b.
\label{vevs}
\end{equation}
In the iso-spin invariant limit, $\alpha_1=\alpha_2$ and
$\beta_1=\beta_2$ while in the SU(3) invariant limit,
$\alpha_1=\alpha_2=\alpha_3$ and $\beta_1=\beta_2=\beta_3$.
The general Lagrangian density which defines our model is
\begin{equation}
{\cal L} = - \frac{1}{2} {\rm Tr}
\left( \partial_\mu M \partial_\mu M^\dagger
\right) - \frac{1}{2} {\rm Tr}
\left( \partial_\mu M^\prime \partial_\mu M^{\prime \dagger} \right)
- V_0 \left( M, M^\prime \right) - V_{SB},
\label{mixingLsMLag}
\end{equation}
where $V_0(M,M^\prime) $ stands for a general function made
from SU(3)$_{\rm L} \times$ SU(3)$_{\rm R}$
(but not necessarily U(1)$_{\rm A}$) invariants
formed out of
$M$ and $M^\prime$. The last term, $V_{SB}$, stands for chiral symmetry
breaking terms which transform in the same way
as the quark mass terms in
the fundamental QCD Lagrangian. In the present paper
we shall, in later sections, specialize to the
zero quark mass limit by taking $V_{SB}=0$.
Not only does this make the formalism much simpler but
it is well known that, due to the spontaneous breakdown of
chiral symmetry, the main mechanism of physical hadron
mass generation is already accounted for.
This is convenient for
disentangling the general properties
of each multiplet from
the uncertainty as to which of the many possible mass type
terms in the efffective Lagrangian to include. We
record the behaviors
of the fields under infinitesimal transformations. Let us write the
infinitesimal vector (L+R) and axial vector (L-R) transformations
of $\phi$ and $S$ as,
\begin{eqnarray}
\delta_V \phi &=& [E_V,\phi], \quad \quad \delta_A \phi=-i[E_A,S]_+,
\nonumber \\
\delta_V S &=& [E_V,S], \quad \quad \delta_A S=i[E_A,\phi]_+.
\label{inftrans}
\end{eqnarray}
Here, unitarity demands that the infinitesimal matrices
obey,
\begin{equation}
E_V^{\dagger}=-E_V, \quad \quad E_A^{\dagger}=-E_A.
\label{unitarity}
\end{equation}
If we demand that the transformations be unimodular, so
that the U(1)$_{\rm A}$ transformation is not included
(the U(1)$_{\rm V}$ transformation is trivial for mesons), we should
also impose ${\rm Tr}(E_A)=0$. However we will not do this so the effects
of
U(1)$_{\rm A}$ will also be included. The transformation properties
of the  $qq{\bar q}{\bar q}$ type fields are:
\begin{eqnarray}
\delta_V\phi' &=& [E_V,\phi'], \quad \quad \delta_A \phi'=-i[E_A,S']_+
+2iS' {\rm Tr} (E_A),
\nonumber \\
\delta_V S' &=& [E_V,S'], \quad \quad \delta_A S'=i[E_A,\phi']_+
-2i\phi' {\rm Tr} (E_A).
\label{inftranspr}
\end{eqnarray}
The extra terms for the axial transformations reflect the different
U(1)$_{\rm A}$ transformation properties of $M$ and $M'$.

    We will employ two complementary approaches to make predictions. One
approach will be to study generating equations for tree level vertices.
These are like Ward identities and follow for any choice of
$V_0(M,M^\prime)$ in Eq. (\ref{mixingLsMLag}).
These predictions are consistent with but will not give all possible
predictions which would arise if one considered, as a second approach,
making a specific choice of
terms in $V_0(M,M^\prime)$.

    The method of treatment, as used earlier \cite{SU1} to discuss the
model
containing only the field $M$, is based on two
generating equations which reflect the invariance of $V_0$ under
vector and axial vector transformations. Differentiating them once,
relates two point vertices (masses) with one point vertices.
Differentiating them twice relates three point vertices
(trilinear couplings) with masses and so on.
Under the infinitesimal vector and axial vector transformations
we have,
\begin{eqnarray}
\delta_V V_0&=&  {\rm Tr} \left(\frac{\partial V_0}{\partial \phi}\delta_V
\phi
             +\frac{\partial V_0}{\partial S}\delta_VS\right)
             +(\phi,S)\rightarrow(\phi',S') =0 ,
    \nonumber \\
\delta_AV_0&=&  {\rm Tr} \left(\frac{\partial V_0}{\partial \phi}\delta_A
\phi
             +\frac{\partial V_0}{\partial S}\delta_AS\right)
             +(\phi,S)\rightarrow(\phi',S') =- {\cal L}_\eta ,
\label{V0invariance}
\end{eqnarray}
wherein the non-zero value of the axial variation equation
reflects the presence in $V_0$  of any terms which are
not invariant under U(1)$_{\rm A}$; these terms will provide mass to the
$\eta^\prime(958)$ meson.
In \cite{SU1}, terms of this type were represented by any function
of the chiral SU(3), but not U(1)$_{\rm A}$, invariant det(M) plus its
hermitian
conjugate. After QCD, 't Hooft found \cite{t} that such a form would arise
from instanton effects. If one wishes to model the  U(1)$_{\rm A}$
anomaly equation of QCD in the single M model the suggested form \cite{ln}
is:
\begin{equation}
{\cal L}_\eta =-c_3\left[{\rm ln} \left(\frac{{\rm det} M}{{\rm det}
M^{\dagger}}\right)\right]^2,
\label{etaprmass}
\end{equation}
where $c_3$ is a numerical parameter.
In the present $M-M'$ model this form is not unique and
the most plausible modification \cite{modlog} is to replace
${\rm ln} (\frac{{\rm det} M}{{\rm det} M^{\dagger}})$ by
\begin{equation}
\gamma_1\,
{\rm ln} \left(
      {
         { {\rm det}(M)}
             \over
        { {\rm det}(M^\dagger)}
      }
\right)
+(1-\gamma_1) \,
{\rm ln}\left(
{
   { {\rm Tr}(MM'^\dagger) }
            \over
   { {\rm Tr}(M'M^\dagger) }
}
\right),
\label{gamma1}
\end{equation}
where $\gamma_1$ is a dimensionless parameter.
Using Eqs. (\ref{inftrans}) and
(\ref{inftranspr})
as well as the arbitrariness of the variations $E_V$ and $E_A$ yields
the matrix
generating equations,
\begin{eqnarray}
&& \left[ \phi,\frac{\partial V_0}{\partial \phi}\right] +
\left[ S,\frac{\partial V_0}{\partial S}\right] +
(\phi,S)\rightarrow(\phi',S')
=0,
\nonumber \\
&& \left[ \phi,\frac{\partial V_0}{\partial S}\right]_+ -
\left[ S,\frac{\partial V_0}{\partial \phi}\right]_+ +
(\phi,S)\rightarrow(\phi',S') =
1\left[2 {\rm Tr}\left(\phi'\frac{\partial V_0}{\partial S'}-
S'\frac{\partial V_0}{\partial \phi'}\right) - 8c_3i {\rm
ln}\left (\frac{{\rm det}
M}{{\rm det}
M^{\dagger}}\right)\right],
\label{geneqs}
\end{eqnarray}
where the form of Eq. (\ref{etaprmass}) was used.
In addition, the replacement, Eq.(\ref{gamma1}) should be borne
in mind.
To get constraints on the particle masses we will differentiate
these equations once with respect to each of the four matrix fields:
$\phi,\phi',S,S'$ and evaluate the equations in the ground state.
Thus we also need the ``minimum" condition,
\begin{equation}
\left\langle \frac{\partial V_0}{\partial S}\right\rangle +
\left\langle
\frac{\partial
V_{SB}}{\partial
S}
\right\rangle
=0,
\quad \quad
\left\langle
\frac{\partial V_0}{\partial S'}
\right\rangle
+
\left\langle
\frac{\partial
V_{SB}}{\partial S'}
\right\rangle
=0.
\label{mincond}
\end{equation}
In ref \cite{FJS05} we considered the canonical term,
$V_{SB}=-2 {\rm Tr} (AS)$ as an effective representation of the
fundamental
quark mass terms; $A$ is a diagonal matrix with entries
proportional to the three
quark masses.
Next, let us differentiate successively the axial vector generating
equation with respect to $\phi$ and to $\phi'$. It is neater to write
the results first for the case when fields with different upper and lower
tensor indices are involved:

\begin{eqnarray}
(\alpha_a + \alpha_b)
\left\langle
{
  {\partial^2 V_0} \over {\partial \phi_b^a \partial \phi_a^b}
}
\right\rangle
+ (\beta_a + \beta_b)
\left\langle
{
  {\partial^2 V_0} \over {\partial {\phi'}_b^a \partial \phi_a^b}
}
\right\rangle
&=& 2(A_a+A_b),
   \nonumber \\
(\alpha_a + \alpha_b)
\left\langle
{
  {\partial^2 V_0} \over {\partial {\phi'}_b^a \partial {\phi}_a^b}
}
\right\rangle
+ (\beta_a + \beta_b)
\left\langle
{
  {\partial^2 V_0} \over {\partial {\phi'}_b^a \partial {\phi'}_a^b}
}
\right\rangle
&=& 0
\label{offdiagpsmasses}
\end{eqnarray}

Next, let us write the corresponding equations
for the case when the upper and lower tensor indices on each field
are the same.
\begin{eqnarray}
\alpha_b
\left\langle
{
  {\partial^2 V_0} \over {\partial \phi_a^a \partial \phi_b^b}
}
\right\rangle
+ \beta_b
\left\langle
{
  {\partial^2 V_0} \over {\partial {\phi}_a^a \partial {\phi'}_b^b}
}
\right\rangle
&=& \sum_g \beta_g
\left\langle
{
  {\partial^2 V_0} \over {\partial {\phi}_a^a \partial {\phi'}_g^g}
}
\right\rangle
-\frac{8c_3}{\alpha_a},
   \nonumber \\
\alpha_b
\left\langle
{
  {\partial^2 V_0} \over {\partial {\phi'}_a^a \partial \phi_b^b}
}
\right\rangle
+ \beta_b
\left\langle
{
  {\partial^2 V_0} \over {\partial {\phi'}_a^a \partial {\phi'}_b^b}
}
\right\rangle
&=& \sum_g \beta_g
\left\langle
{
  {\partial^2 V_0} \over {\partial {\phi'}_a^a \partial {\phi'}_g^g}
}
\right\rangle.
\label{pscalarmasses}
\end{eqnarray}
Note that the axial generating equation provides information on
the masses of all the pseudoscalars. Further differentiations
will relate a large number of trilinear and quadrilinear
coupling constants to the meson masses and to the quark mass
coefficients, $A_a$.

To fully characterize the system we will also require some
knowledge of the axial vector and vector currents \cite{SU1} obtained by
Noether's method:
\begin{eqnarray}
(J_\mu^{axial})_a^b &=&(\alpha_a+\alpha_b)\partial_\mu\phi_a^b +
(\beta_a+\beta_b)\partial_\mu{\phi'}_a^b+ \cdots,
\nonumber \\
(J_\mu^{vector})_a^b &=&i(\alpha_a-\alpha_b){\partial_\mu} S_a^b +
i(\beta_a-\beta_b)\partial_\mu {S'}_a^b+ \cdots,
\label{currents}
\end{eqnarray}
where the dots stand for terms bilinear in the fields.

It will be helpful to briefly review
the treatment of the $\pi$-$\pi^\prime$ system as given in section
IV of ref. \cite{FJS05}. Introduce the abbreviations
\begin{eqnarray}
x_\pi &=& \frac{2A_1}{\alpha_1},
\nonumber \\
y_\pi &=&\left\langle \frac{\partial^2V}{\partial
{\phi'}_2^1\partial{\phi'}_1^2}
\right\rangle,
\nonumber \\
z_\pi&=& \frac{\beta_1}{\alpha_1}.
\label{xyzpi}
\end{eqnarray}
Here we have introduced the total potential $V=V_0+V_{SB}$.
Substituting $a=1, b=2$ into
both of Eqs. (\ref{offdiagpsmasses}) enables us to
write the (non-diagonal) matrix of squared $\pi$
and $\pi'$ masses as:
\begin{equation}
(M_\pi^2)=\left[ \begin{array}{c c}
                x_\pi +z_\pi^2 y_\pi & -z_\pi y_\pi
\nonumber \\
                -z_\pi y_\pi   & y_\pi
                \end{array} \right] .
\label{Mpi}
\end{equation}
We see that
$x_\pi$ would be the squared pion mass in the single $M$
model and $y_\pi$ represents the squared mass of
the ``bare" $\pi'$.
  The transformation between the diagonal
fields (say $\pi^+$ and $\pi'^+$)  and the
original pion fields is defined as:
\begin{equation}
\left[
\begin{array}{c}  \pi^+ \\
                 \pi'^+
\end{array}
\right]
=
R_\pi^{-1}
\left[
\begin{array}{c}
                        \phi_1^2 \\
                        {\phi'}_1^2
\end{array}
\right]=
\left[
\begin{array}{c c}
                \cos\theta_\pi & - \sin \theta_\pi
\nonumber               \\
\sin \theta_\pi & \cos \theta_\pi
\end{array}
\right]
\left[
\begin{array}{c}
                        \phi_1^2 \\
                        {\phi'}_1^2
\end{array}
\right],
\label{mixingangle}
\end{equation}
which also defines the transformation matrix, $R$.
The explicit diagonalization gives an expression for the
mixing angle $\theta_\pi$:
\begin{equation}
{\rm tan} (2\theta_\pi)=\frac{-2y_\pi z_\pi}{y_\pi(1-z_\pi^2)-x_\pi}.
\label{thetasubpi}
\end{equation}
The mixing angle, $\theta_\pi$ can be connected to the
experimentally known value of the pion decay constant.
Substituting the expressions from Eq. (\ref{mixingangle}) for $\phi_1^2$
and ${\phi'}_1^2$ in terms of the physical fields $\pi^+$
and $\pi'^+$ into Eq. (\ref{currents}) yields,
\begin{eqnarray}
(J_\mu^{axial})_1^2 &=&F_\pi\partial_\mu \pi^+ + F_{\pi'}\partial_\mu
\pi'^+
+\cdots,
\nonumber \\
F_\pi &=&(\alpha_1+\alpha_2) \cos\theta_\pi -
(\beta_1+\beta_2)\sin\theta_\pi,
\nonumber \\
F_{\pi'} &=&(\alpha_1+\alpha_2)\sin\theta_\pi +
(\beta_1+\beta_2)\cos\theta_\pi.
\label{Fpis}
\end{eqnarray}

\section{Simplification for zero quark masses}
    The zero quark mass limit is gotten by taking $V_{SB}=0$. We assume
that
the original SU(3)$_{\rm L} \times$ SU(3)$_{\rm R}$ symmetry is
spontaneously broken to
SU(3)$_{\rm V}$ rather than some smaller subgroup. The vacuum expectation
values
of the scalar fields simplify to:
\begin{equation}
\alpha_1=\alpha_2=\alpha_3=\alpha,\hspace{1cm}\beta_1=\beta_2=\beta_3
=\beta.
\label{alphabeta}
\end{equation}
The mass spectrum also simplifies a lot. When quark masses
are included
in the isotopic spin invariant approximation there
are 16 different
masses. However in the zero quark mass limit there are only 8
different masses. These describe the four systems of
degenerate SU(3) octet or SU(3) singlet fields:
\begin{eqnarray}
({\hat \phi},{\hat \phi^\prime}),\hspace{1cm}(\phi_0,\phi_0^\prime),
\nonumber \\
({\hat S},{\hat S^\prime}),\hspace{1cm}(S_0,S_0^\prime).
\label{4systems}
\end{eqnarray}
Here the hat stands for the eight members of the appropriate
octets.
The fields of each system can mix with each other but not with
the fields of any other system. In addition to 8 different masses
there will be four different mixing angles describing four orthogonal
2$\times$2 matrices. The conventions are the same as
in Eq. (\ref{mixingangle}) so that $\theta_{\pi}$ now describes
the mixings of the two pseudoscalar octets. Note that if only isotopic
spin invariance were present, the isotopic spin zero fields
of each parity would
be characterized by a 4$\times$4 mixing matrix with 6 angle parameters
(See Eq.(64) of \cite{FJS05} for example).
Notice also that the $\pi-\pi^\prime$, $K-K^\prime$ and
$\eta_8-\eta_8^\prime$ mixings,
for example, are all described by the same mixing parameter
$\theta_{\pi}$.

    We next discuss the notations for resolving the
nonets
into SU(3) octets and singlets. Matrix notation is sometimes convenient;
we use the convention $\phi^b_a\rightarrow\phi_{ab}$. The properly
normalized singlet states are:
\begin{eqnarray}
\phi_0=\frac{1}{\sqrt{3}}{\rm Tr}(\phi),\hspace{1cm}
\phi_0^\prime=\frac{1}{\sqrt{3}}{\rm Tr}(\phi^\prime),
\nonumber \\
S_0=\frac{1}{\sqrt{3}}{\rm Tr}(S),\hspace{1cm}
S_0^\prime=\frac{1}{\sqrt{3}}{\rm Tr}(S^\prime).
\label{singlets}
\end{eqnarray}
Then we have the matrix decompositions:
\begin{eqnarray}
\phi={\hat \phi}+\frac{1}{\sqrt{3}}\phi_{0}1,\hspace{1cm}
\phi^\prime={\hat \phi^\prime}+\frac{1}{\sqrt{3}}\phi_{0}^\prime1,
\nonumber \\
S={\hat S}+\frac{1}{\sqrt{3}}S_{0}1,\hspace{1cm}
S^\prime={\hat S^\prime}+\frac{1}{\sqrt{3}}S_{0}^{\prime}1,
\label{octets}
\end{eqnarray}
wherein ${\hat \phi}$, ${\hat \phi^\prime}$, ${\hat S}$ and
${\hat S^\prime}$
are all 3$\times$3 traceless matrices.
The singlet scalar fields may be further
decomposed as:
\begin{equation}
S_0=\sqrt{3}\alpha+{\tilde S_0},\hspace{1cm}
S_0^\prime=\sqrt{3}\beta+{\tilde S_0}^\prime.
\label{scalarsinglets}
\end{equation}
Here ${\tilde S_0}$ and ${\tilde S_0}^\prime$ are the
fluctuation fields around the true ground state of the model.

    Setting $x_\pi=0$, corresponding to zero quark masses,
simplifies Eq. (\ref{thetasubpi}) for the $\pi$-$\pi^\prime$
mixing angle to:
\begin{equation}
{\rm tan} {2\theta_\pi}=\frac{-2z_\pi}{1-z_\pi^2}\equiv
\frac{2 {\rm tan} {\theta_\pi}}{1- {\rm tan}^2{\theta_\pi}}.
\label{2thetapi}
\end{equation}
This immediately yields:
\begin{equation}
{\rm tan}{\theta_\pi}=-\frac{\beta}{\alpha}.
\label{thetapi}
\end{equation}
Substituting this into Eq. (\ref{Fpis}) yields the simple results:
\begin{eqnarray}
F_\pi&=&2\sqrt{\alpha^2+\beta^2},
\nonumber \\
F_{\pi^\prime} &=&0.
\label{nomassFpis}
\end{eqnarray}

    One may note, for comparison, from Table 4 in \cite{FJS052}
that $F_{\pi^\prime}$ and also $F_{K^\prime}$ are not exactly zero
in the presence of non zero quark masses, although they
are very heavily
suppressed. This feature suggests the essential reliability of the
zero quark mass limit.

    Next consider the pseudoscalar octets, ${\hat \phi}$
and ${\hat \phi^\prime}$ in the model. Because of SU(3) symmetry it
is sufficient
to give just the two $I=I_3=1$ fields, $\phi_1^2$ and ${\phi'}_1^2$.
Their mixing matrix, Eq. (\ref{Mpi}) becomes in the limit of zero quark
masses:
\begin{equation}
(M_\pi^2)=y_\pi\left[ \begin{array}{c c}
                z_\pi^2  & -z_\pi
\nonumber \\
                -z_\pi   & 1
                \end{array} \right] =
\left\langle
{
  {\partial^2 V_0} \over {\partial {\phi'}_1^2 \partial {\phi'}_2^1}
}
\right\rangle
\left[ \begin{array}{c c}
                \beta^2/\alpha^2  & -\beta/\alpha
\nonumber \\
                -\beta/\alpha  & 1
                \end{array} \right]
\label{mpipiprime}
\end{equation}
It is easy to see that this matrix has zero determinant and to identify
the usual (but zero mass) pseudoscalar pion as
\begin{equation}
\pi^+ = \frac{2}{F_\pi}
\left(\alpha\phi^2_1+\beta{\phi'}^2_1\right),
\label{nomasspion}
\end{equation}
where $F_\pi=$131 MeV. The physical massive pion ``excitation" is clearly
$\pi'^+ = \frac{2}{F_\pi}(-\beta\phi^2_1+\alpha{\phi'}^2_1)$ and has a
squared mass, $m^2(\pi')=y_\pi(1+\beta^2/\alpha^2)$. We notice that,
just from our general treatment, the $\pi-\pi'$ system
can be described by the three
parameters $\alpha$, $\beta$ and $y_\pi$. However, there are only two
physical quantities, $F_\pi$ and $m^2(\pi')$, to compare with. Thus
the mixing angle between the usual and the ``excited" pseudoscalar
octet states is not predicted in general. In order to predict this
interesting quantity
we have to specify our  choice of chiral invariant terms
in the potential, $V$. A similar situation will be seen to hold
for trilinear and quadrilinear coupling constants involving the
physical pseudoscalars. There are many constraints just from chiral
symmetry but a complete (though clearly model dependent) description
will depend on the particular choice of terms in the potential.

    It is also amusing to look at the $\phi_0-{\phi'}_0$ sector in
the zero quark mass limit. There is a rather drastic simplification
since the introduction of quark masses results in additional mixing with
the isoscalar members of the corresponding octets. That requires
a six parameter 4$\times$4 transformation matrix rather than the single
parameter 2$\times$2 matrix we now will get. Using the formula,
\begin{equation}
\frac{\partial}{\partial \phi_0}=\frac{1}{\sqrt{3}}
\left(\frac{\partial}{\partial\phi_1^1}+\frac{\partial}{\partial\phi_2^2}
+\frac{\partial}{\partial\phi_3^3}\right),
\nonumber
\end{equation}
in both of Eqs. (\ref{pscalarmasses}), we end up with the
pre-diagonal $\phi_0-{\phi'}_0$ mass squared matrix:
\begin{equation}
(M^2_0)=
\left[ \begin{array}{c c}
z_0^2 y_0
                -\frac{8c_3(2\gamma_1+1)^2}{3\alpha^2}
& -z_0
y_0+\frac{8c_3(1-\gamma_1)(2\gamma_1+1)}{3\alpha\beta}
\nonumber \\
                -z_0 y_0
+\frac{8c_3(1-\gamma_1)(2\gamma_1+1)}{3\alpha\beta}  &
y_0-\frac{8c_3(1-\gamma_1)^2}{3\beta^2}
                \end{array} \right] .
\label{phizeromixing}
\end{equation}
Here $z_0=-2\beta/\alpha$ and
\begin{equation}
y_0=\left\langle \frac{\partial^2
V}{\partial\phi'_0\partial{\phi'}_0}\right\rangle.
\nonumber
\end{equation}
The mixing angle, $\theta_0$, is defined by the convention:
\begin{equation}
\left[
\begin{array}{c}  \phi_{0p} \\
                 \phi'_{0p}
\end{array}
\right]
=
R^{-1}_0
\left[
\begin{array}{c}
                        \phi_0 \\
                        {\phi'}_0
\end{array}
\right]=
\left[
\begin{array}{c c}
                \cos\theta_0 & -\sin \theta_0
\nonumber               \\
\sin \theta_0 & \cos \theta_0
\end{array}
\right]
\left[
\begin{array}{c}
                        \phi_0 \\
                        {\phi'}_0
\end{array}
\right],
\label{phizeromixingangle}
\end{equation}
    In the limit where $c_3$, defined in Eq. (\ref{etaprmass}), vanishes it 
is
seen that the determinant of the mass squared matrix in Eq.
(\ref{phizeromixing}) vanishes. This is understandable since
$c_3$ multiplies the terms which violate U(1)$_{\rm A}$
symmetry and a zero mass singlet pseudoscalar boson must exist since
the symmetry is broken spontaneously. In this limit the mixing angle
is related to the pseudoscalar octet one by
$\tan{\theta_0}=-2\tan{\theta_\pi}$. It should be remarked that the
effect of non zero $c_3$ is actually quite large so the limit where
it vanishes is mainly of academic interest.

    Of course, the U(1)$_{\rm A}$ transformation is relevant in setting up
this model since, as seen from Eq. (\ref{MU1A}), it distinguishes the two
quark fields from the four quark fields. We shall consider here, models
in which the terms multiplied by $c_3$ are the only ones which violate
U(1)$_{\rm A}$ symmetry. In that case the divergence of the axial current
in the
model exactly mocks up the QCD axial anomaly at tree level. Alternatively,
a term like ${\rm det} (M) +h.c.$ could be used with similar results; such
a term
does not however mock up the U(1)$_{\rm A}$ anomaly equation.

    We have seen that quite a lot of information about the pseudoscalar
particle masses and mixings follows just from the axial generating
equations, reflecting the spontaneous breakdown of the octet axial
symmetries. On the
other hand, Eq.(31) of ref.\cite{FJS05} shows that, in the case
where spontaneous breakdown preserves the SU(3) invariance of the vacuum,
there will be no such model independent
information about the masses and
mixings of the scalars. To find that information, one must make models
with specific
choices of the invariant terms. In preparation we give
notations for the scalar mass and transformation matrices, analogous
to those we adopted for the pseudoscalars, in the case where quark masses
are absent and the vacuum is assumed to be SU(3)$_{\rm V}$ invariant.
The pre-diagonal 2 $\times$ 2 matrix for the $I=I_3=1$ scalar meson
squared
masses
is denoted
$(X_a^2)$ and  the mass diagonal fields, $a^+$ and $a'^+$ are related
to the non-diagonal ones by:
\begin{equation}
\left[
\begin{array}{c}  a^+ \\
                 a'^+
\end{array}
\right]
=
L_a^{-1}
\left[
\begin{array}{c}
                        S_1^2 \\
                        {S'}_1^2
\end{array}
\right]=
\left[
\begin{array}{c c}
                \cos\psi_a & -\sin \psi_a
\nonumber               \\
\sin \psi_a & \cos \psi_a
\end{array}
\right]
\left[
\begin{array}{c}
                        S_1^2 \\
                        {S'}_1^2
\end{array}
\right].
\label{amixingangle}
\end{equation}
This is sufficient to describe the mixing of all the scalar octet
particles with corresponding SU(3) quantum numbers. For
the $S_0-S'_0$ mixing, we define the prediagonal squared mass
matrix to be
$(X_0^2)$ while the mass diagonal fields $S_{0p}$ and $S'_{0p}$ are
defined by:
\begin{equation}
\left[
\begin{array}{c}  S_{0p} \\
                 S'_{0p}
\end{array}
\right]
=
L_0^{-1}
\left[
\begin{array}{c}
                        S_0 \\
                        S'_0
\end{array}
\right]=
\left[
\begin{array}{c c}
                \cos \psi_0 & -\sin \psi_0
\nonumber               \\
\sin \psi_0 & \cos \psi_0
\end{array}
\right]
\left[
\begin{array}{c}
                        S_0 \\
                        S'_0
\end{array}
\right],
\label{szeromixingangle}
\end{equation}

\section{Model for masses and mixings}
    As just discussed, it is necessary to make a specific choice of terms
in the SU(3)$_L\times$ SU(3)$_R$ invariant potential $V_0$ in order to be
able to predict all physical properties of the system. This is a
non trivial issue
since, for example, if we restrict $V_0$ to be renormalizable, there are
twenty one terms \cite{FJS05} with this symmetry. We will adopt
two criteria for which terms to include.
First we list the six SU(3)$_L\times$ SU(3)$_R$
invariant terms which satisfy these criteria and seem
the most reasonable for an initial treatment:
\begin{eqnarray}
V_0 =&-&c_2 \, {\rm Tr} (MM^{\dagger}) +
c_4^a \, {\rm Tr} (MM^{\dagger}MM^{\dagger})
\nonumber \\
&+& d_2 \,
{\rm Tr} (M^{\prime}M^{\prime\dagger})
     + e_3^a \left(\epsilon_{abc}\epsilon^{def}M^a_dM^b_eM'^c_f +
h.c.\right)
\nonumber \\
     &+&  c_3 \left[ \gamma_1 {\rm ln} \left(\frac{{\rm det} M}{{\rm det}
M^{\dagger}}\right)
+(1-\gamma_1){\rm ln}
\left(\frac{{\rm Tr}(MM'^\dagger)}{{\rm
Tr}(M'M^\dagger)}\right)\right]^2.
\label{SpecLag}
\end{eqnarray}
     All the terms except the last two have been chosen to also
possess the  U(1)$_{\rm A}$
invariance.
Those terms are clearly  non-renormalizable and violate U(1)$_{\rm A}$
invariance in a special way.
They have, as previously discussed [see Eq.(\ref{gamma1})], the correct
U(1)$_{\rm A}$
property so
that
the resulting Lagrangian can exactly mock up the U(1)$_{\rm A}$ anomaly of
QCD.
Of course, we
are using the effective Lagrangian at tree level and renormalizability
is not an issue at this level.
  Renormalizable terms of the instanton
determinant type and the type ${\rm Tr}(MM'^\dagger)+h.c.$
could be used instead with not much change in the result. However,
the
role that the U(1)$_{\rm A}$ transformation is playing in distinguishing
``four
quark" from ``two quark" effective fields suggests
that we try to reproduce as much as possible of the behavior of QCD
under axial U(1)$_{\rm A}$. The ln terms chosen also have the convenient
feature that they confine the U(1)$_{\rm A}$ violating effects to the
SU(3) singlet pseudoscalar sector of the model.
The first four  terms were chosen
from the twelve renormalizable and U(1)$_{\rm A}$ invariant ones in the
formula,
Eq. (A1) of \cite{FJS05} (please see also Appendix A of the present paper)
by imposing
the criterion that effective vertices describing the smallest
numbers of quarks plus antiquarks be retained. This quantity,
representing the total number of fermion lines at each
effective vertex can
be written as,
\begin{equation}
N=2n+4n^\prime,
\label{N}
\end{equation}
where $n$ is the number of times $M$ or $M^\dagger$ appears
in each term while
$n^\prime$ is the number of times $M^\prime$ or ${M^\prime}^\dagger$
appears in each term..
Thus, the $c_2$ term has N=4  while the
$c_4^a$, $d_2$ and $e_3^a$ terms each have N=8.
For simplicity, we have neglected the $N$=8 term,
$c_4^b[{\rm Tr}(MM^\dagger)]^2$ which is suppressed, in the single $M$
model, by the quark line rule.
It may be noted that the quantities ${\rm det}(M)$ and ${\rm
Tr}(MM'^\dagger)$
which enter into those two terms which saturate the
U(1)$_{\rm A}$ anomaly have $N$=6.
On the other hand, the terms in
\begin{equation}
      e_4^a \, {\rm Tr} (MM^{\dagger}M^{\prime}M^{{\prime}{\dagger}})
      + e_4^b \, {\rm Tr} (MM^{{\prime}{\dagger}}M^{\prime}M^{\dagger})
\label{e4}
\end{equation}
each represent twelve quarks plus antiquarks at the same vertex and
will not be included at the present stage. Similarly, the term
$d_4^a {\rm Tr}(M'M'^{\dagger}M'M'^{\dagger})$ representing sixteen quarks
and antiquarks will not be included. In the future, U(1)$_{\rm A}$
invariant
terms with higher values of $N$ may be used to systematically
improve the approximation as well as U(1)$_{\rm A}$
violating operators with
higher values of $N$ which may be inserted into an obvious generalization
of Eq.(\ref{gamma1}).
    The minimum equations for this potential are:
\begin{equation}
\left\langle { {\partial V_0} \over {\partial S_a^a} } \right\rangle =
2 \,\alpha\,  \left( - c_2 + 2\, c_4^a\, \alpha^2 + 4\, e_3^a \, \beta
\right) =0,
\label{mealpha}
\end{equation}

\begin{equation}
\left\langle { {\partial V_0} \over
{\partial {S'}_a^a} } \right\rangle
=
2  \left( d_2\, \beta + 2\, e_3^a\, \alpha^2 \right) = 0.
\label{mebeta}
\end{equation}

      Notice that $\alpha$ is an overall factor in Eq. (\ref{mealpha})
so that, in addition to the physical spontaneous breakdown solution where
$\alpha \ne 0$ there is a solution with $\alpha= 0$. On the other
hand, $\beta$ is not an overall factor of Eq. (\ref{mebeta}) and it is
easy to see that $\beta$ is necessarily non-zero in the physical
situation where $\alpha$ is non-zero. The minimum equations clearly
eliminate two parameters from the model.

   Next, we shall give the matrix elements of the four squared mass mixing
matrices based on the use of the specific potential of
Eq.(\ref{SpecLag}). First consider the matrix describing any of the eight
degenerate $0^-$ quark-antiquark fields mixing with their corresponding
four quark partners. Without using the minimum equations, one obtains:
\begin{equation}
(M^2_\pi) =
\left[
\begin{array}{cc}
2 \, \left(- c_2 + 2\, c_4^a\, \alpha^2 + 2\, e_3^a \, \beta \right) &
4 \, e_3^a\, \alpha \\
4 \, e_3^a\, \alpha &
2d_2
\end{array}
\right]
\label{rawmpisq}
\end{equation}
This corresponds to the general form given in Eq.(\ref{mpipiprime}) when
we
identify,
$y_\pi=2d_2$ and $z_\pi=-2\alpha e_3^a/d_2$. Note that $z_\pi\equiv
\beta/\alpha$.

    The matrix describing the mixing of the two pseudoscalar singlets is
similarly written as:
\begin{equation}
(M^2_0) =
\left[
\begin{array}{cc}
-2 \, \left( c_2 - 2\, c_4^a\,
\alpha^2 + 4\, e_3^a \, \beta
\right)
-\frac{8c_3(2\gamma_1+1)^2}{3\alpha^2}
&
-8  \, e_3^a\, \alpha
+\frac{8c_3(1-\gamma_1)(2\gamma_1+1)}{3\alpha\beta}\\
-8  \, e_3^a\, \alpha +\frac{8c_3(1-\gamma_1)(2\gamma_1+1)}{3\alpha\beta}&
2d_2 -\frac{8c_3(1-\gamma_1)^2}{3\beta^2}
\end{array}
\right]
\label{rawm0sq}
\end{equation}
This corrsponds to the general form given in Eq.(\ref{mpipiprime}) when we
identify, $y_0=2d_2$ and
$z_0=-2\beta/\alpha=4e_3^a\alpha/d_2$.

    For the mixing matrix of the octet scalars, the specific potential
of Eq.(\ref{SpecLag}) directly gives:
\begin{equation}
(X_a^2) =
\left[
\begin{array}{cc}
2 \, \left(- c_2 + 6\, c_4^a\, \alpha^2 - 2\, e_3^a \, \beta \right) &
-4\alpha e_3^a  \\
-4\alpha e_3^a  &
2d_2
\end{array}
\right].
\label{s8mass}
\end{equation}
Finally the squared mass mixing matrix for the singlet scalars is
similarly obtained as:
\begin{equation}
(X_0^2) =
\left[
\begin{array}{cc}
2 \, \left(- c_2 + 6\, c_4^a\, \alpha^2 + 4\, e_3^a \, \beta \right) &
8\alpha e_3^a  \\
8\alpha e_3^a  &
2d_2
\end{array}
\right].
\label{s0mass}
\end{equation}

\begin{figure}[t]
\begin{center}
\vskip 1cm
\epsfxsize = 7.5cm
\ \epsfbox{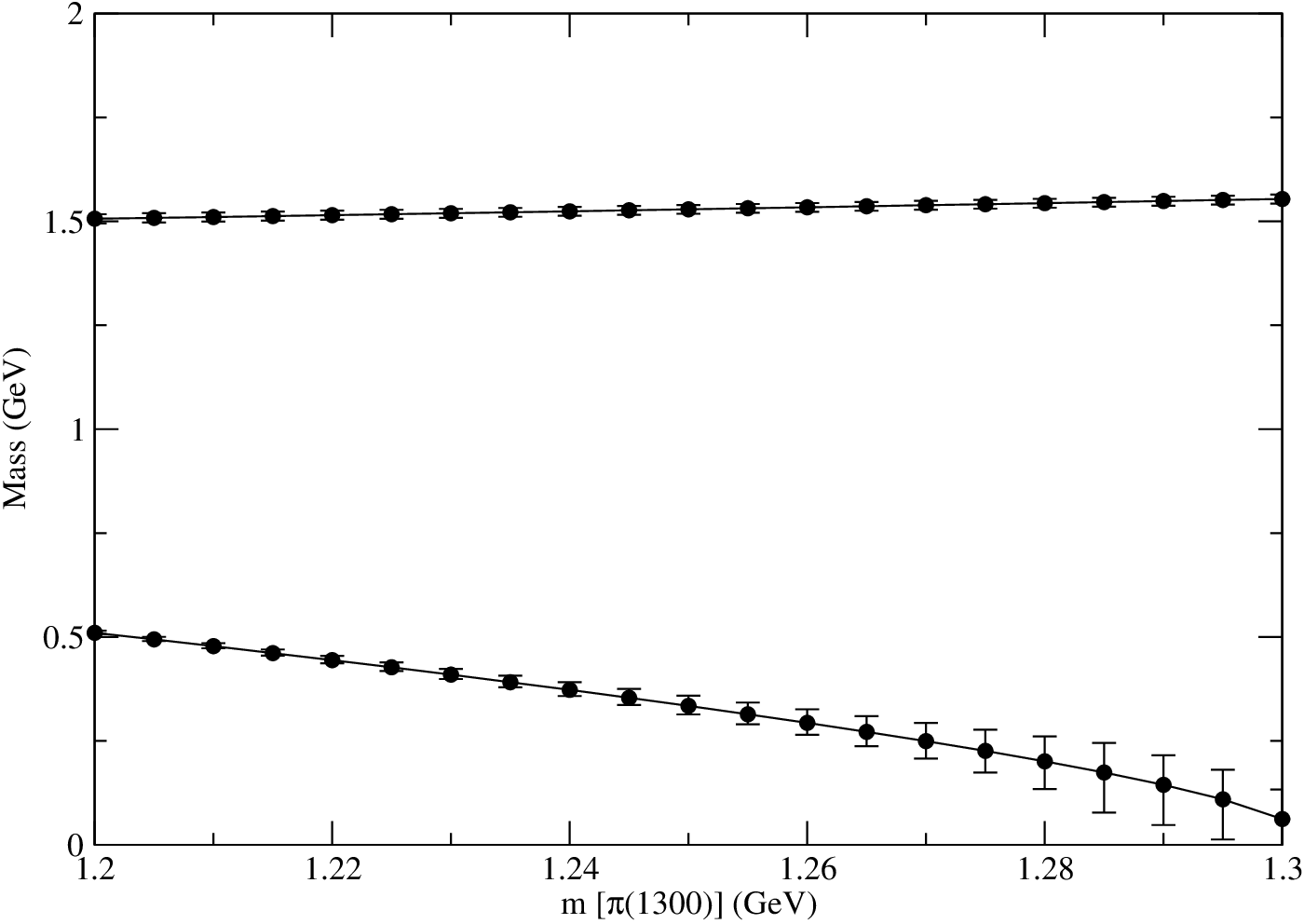}
\end{center}
\caption[]{%
The predictions for the masses of the
two SU(3) singlet scalars
vs. $m[\pi(1300)]$.
The error bars give the effect of the
uncertainty in the $a_0(1450)$ mass.
}
\label{ms0vsmpip}
\end{figure}

     Now let us consider the comparison of this model with experiment.
To start with there are 8 parameters ($\alpha$, $\beta$, $c_2$, $d_2$,
$c_4^a$, $e_3^a$, $c_3$ and $\gamma_1$). These can be reduced to six
by use of the two minimum equations just given. We note that
the parameters  $c_3$ and $\gamma_1$), associated with modeling the
U(1)$_{\rm A}$
anomaly, do not contribute to either the minimum equations or to the
mass matrices of the particles which are not $0^-$ singlets. Thus it is
convenient to first determine the other four independent parameters.
As the corresponding four experimental inputs
\cite{ropp} we take the non-strange
quantities:
\begin{eqnarray}
m(0^+ {\rm octet}) &=& m[a_0(980)] = 984.7 \pm 1.2\, {\rm MeV}
\nonumber \\
m(0^+ {\rm octet}') &=& m[a_0(1450)] = 1474 \pm 19\, {\rm MeV}
\nonumber \\
m(0^- {\rm octet}') &=& m[\pi(1300)] = 1300 \pm 100\, {\rm MeV}
\nonumber \\
F_\pi &=& 131 \, {\rm MeV}
\label{inputs1}
\end{eqnarray}
Evidently, a large experimental uncertainty
appears in the mass of $\pi(1300)$; we shall initially take the other
masses as fixed at their central values and vary this mass in the
indicated range. As shown in Eq.(\ref{lagpara}) in Appendix B, it is
straightforward to determine the four independent parameters
in terms of these masses. There is a complication which must be
taken into account; from
studying the predicted masses of the $0^+$ SU(3) singlet states
one finds that
the  positivity of the eigenvalues
of their squared mass matrix, Eq.(\ref{s0mass}) is only satisfied when,
\begin{equation}
m[\pi(1300)]< 1302\,{\rm MeV}.
\label{scalarconscond}
\end{equation}
Further restrictions on the allowed range of
$m[\pi(1300)]$ will arise when we calculate the
masses of the $0^-$ SU(3) singlet states. Before that we
mention the two predicted masses for the $0^+$ SU(3)
singlet states;
as $m[\pi(1300)]$ varies from 1200 to 1300 MeV,
\begin{eqnarray}
m(0^+ {\rm singlet})&=& 510 \to
28\hspace{.1cm}(410)\hspace{.1cm} {\rm MeV},
\nonumber \\
m(0^+{\rm singlet}') &=&1506
\to 1555\hspace{.1cm}(1520)\hspace{.1cm} {\rm MeV}.
\label{scalsing}
\end{eqnarray}
The predictions in parentheses correspond to the likely additional
constraints from the positivity of the $0^-$ SU(3) singlet states.
Plots are shown in Fig. \ref{ms0vsmpip}.

Clearly, the most dramatic feature is the very low mass of
the lighter SU(3) singlet scalar meson. Of course, one expects the
addition
of quark mass type terms to modify the details somewhat.
On the other hand, there are a number of allowed different quark mass
terms so it is notable that the characteristic very light mass scalar
exists apart from the ambiguity in choice of the quark mass terms.

\begin{figure}[t]
\begin{center}
\epsfxsize = 7.5cm
\epsfbox{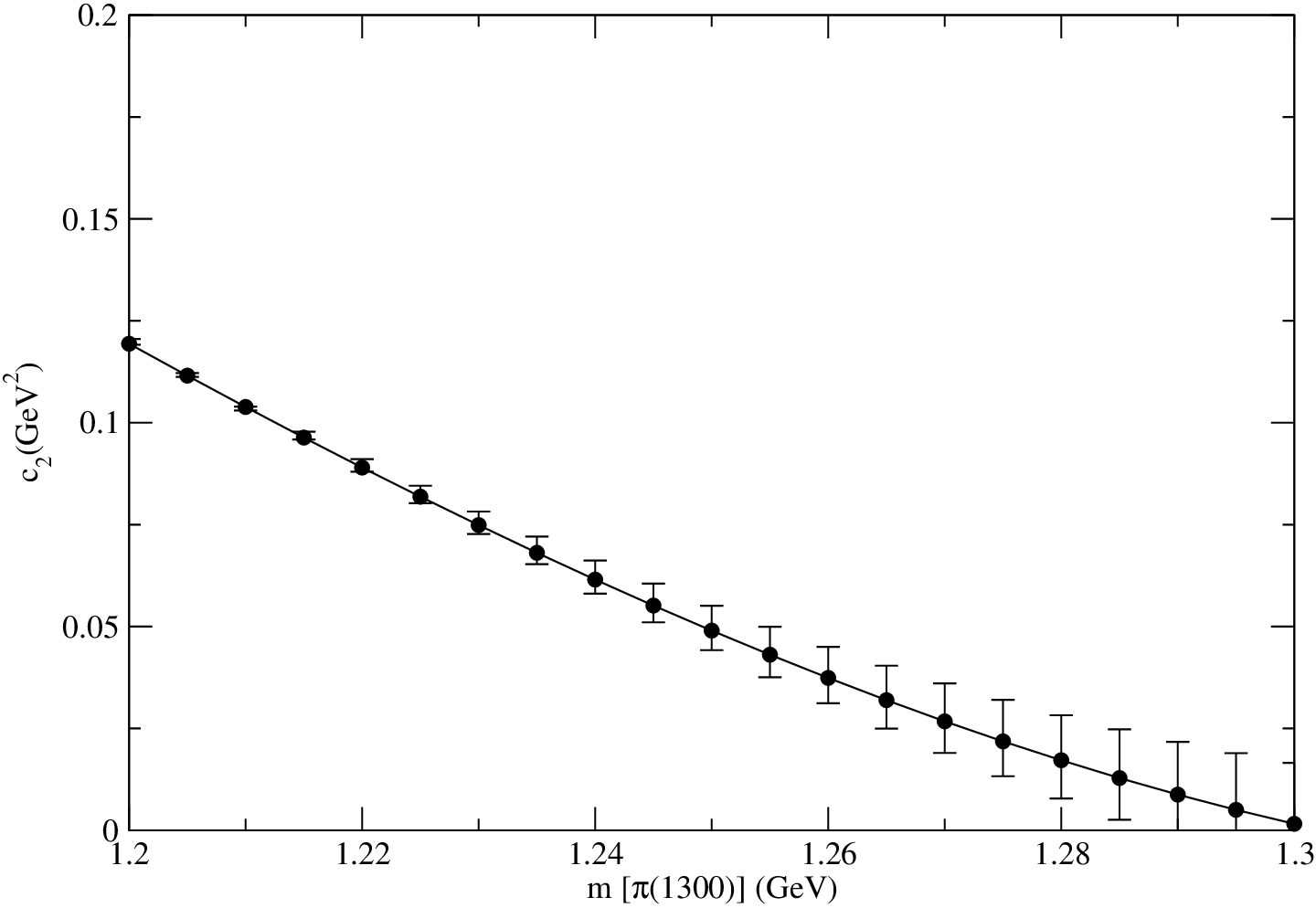}
\hskip 1cm
\epsfxsize = 7.5cm
\epsfbox{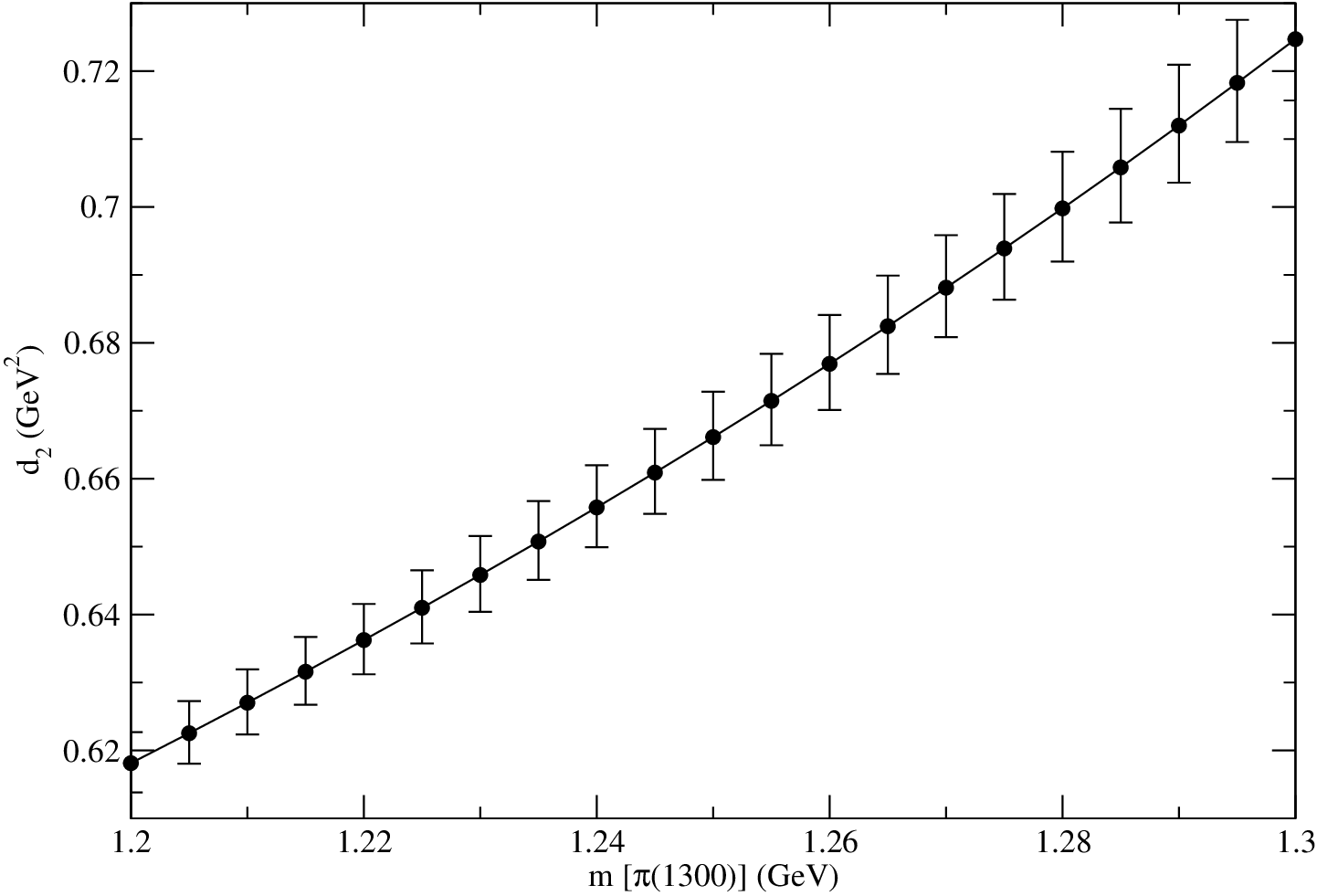}\\

\vskip 1cm

\epsfxsize = 7.5cm
\epsfbox{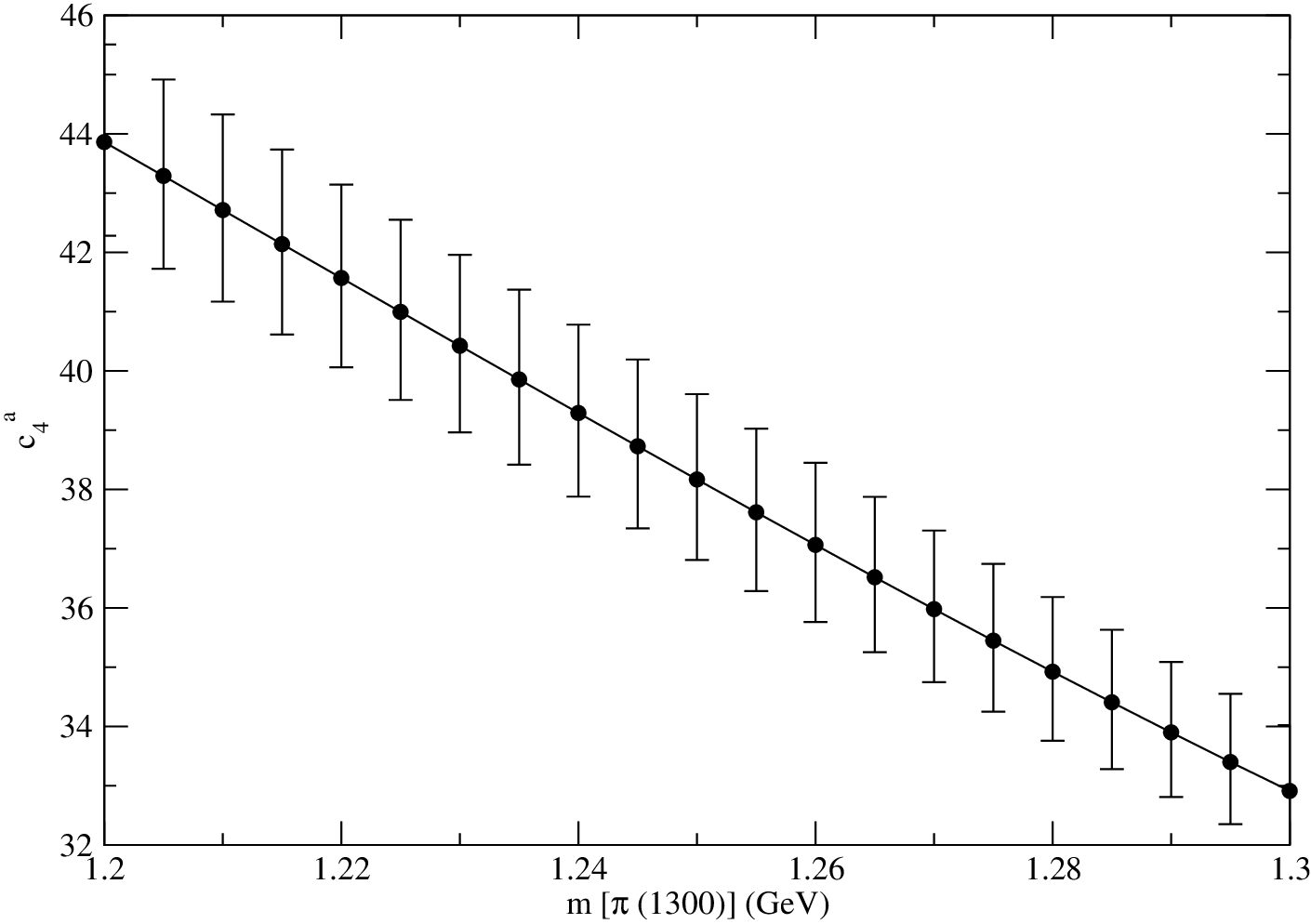}
\hskip 1cm
\epsfxsize = 7.5cm
\epsfbox{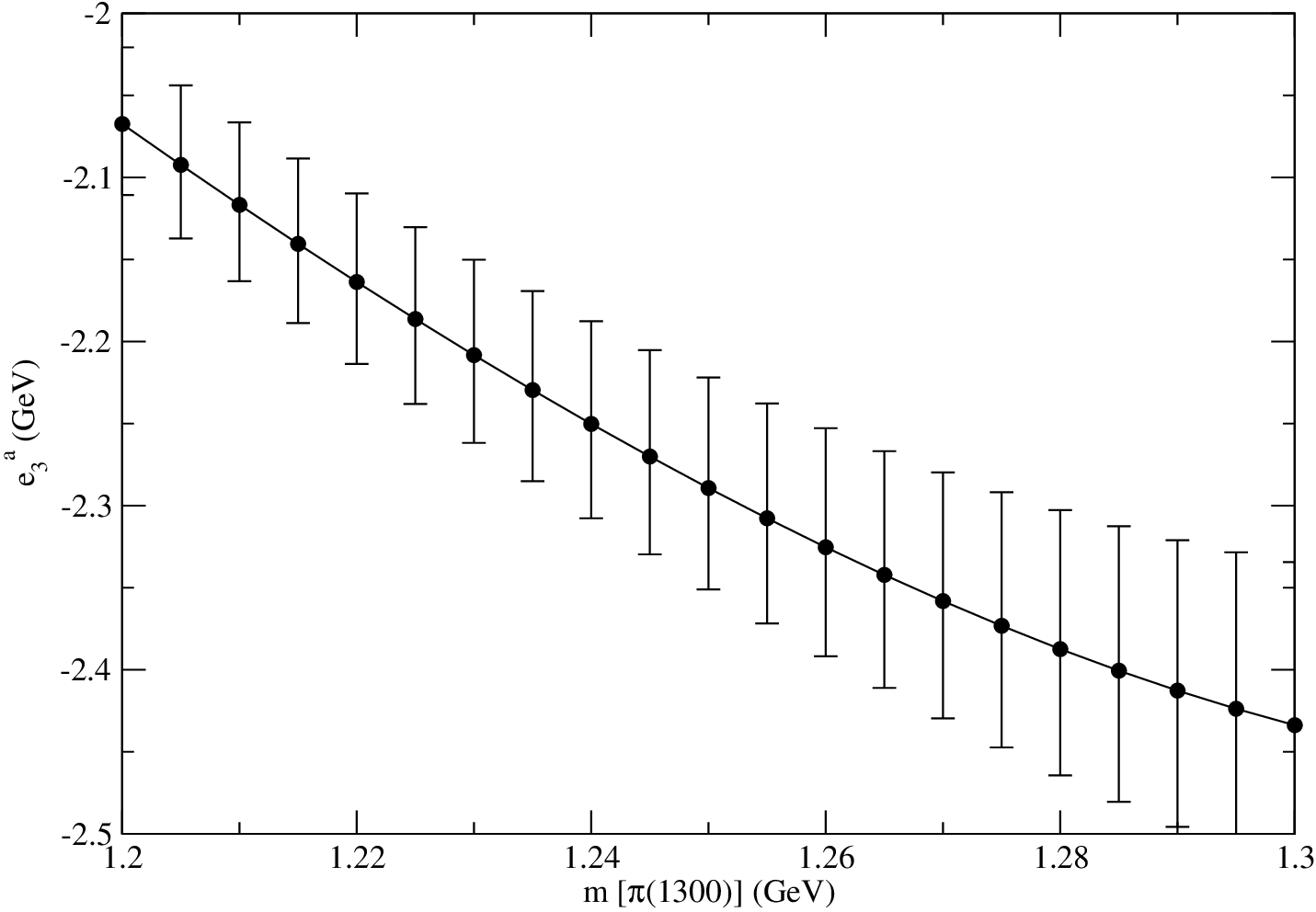}
\end{center}
\caption[]{%
Starting from the upper left and proceeding clockwise:
$c_2$ vs. $m[\pi(1300)]$, $d_2$ vs. $m[\pi(1300)]$, $e_3^a$ vs.
$m[\pi(1300)]$ and  $c_4^a$ vs. $m[\pi(1300)]$. The range of
$m[\pi(1300)]$ corresponds to the restrictions imposed by the positivity
of the scalar SU(3) singlet masses. The error bars give the effect of the
uncertainty in the $a_0(1450)$ mass.
}
\label{ind4}
\end{figure}

\begin{figure}[h]
\begin{center}
\epsfxsize = 7.5cm
\epsfbox{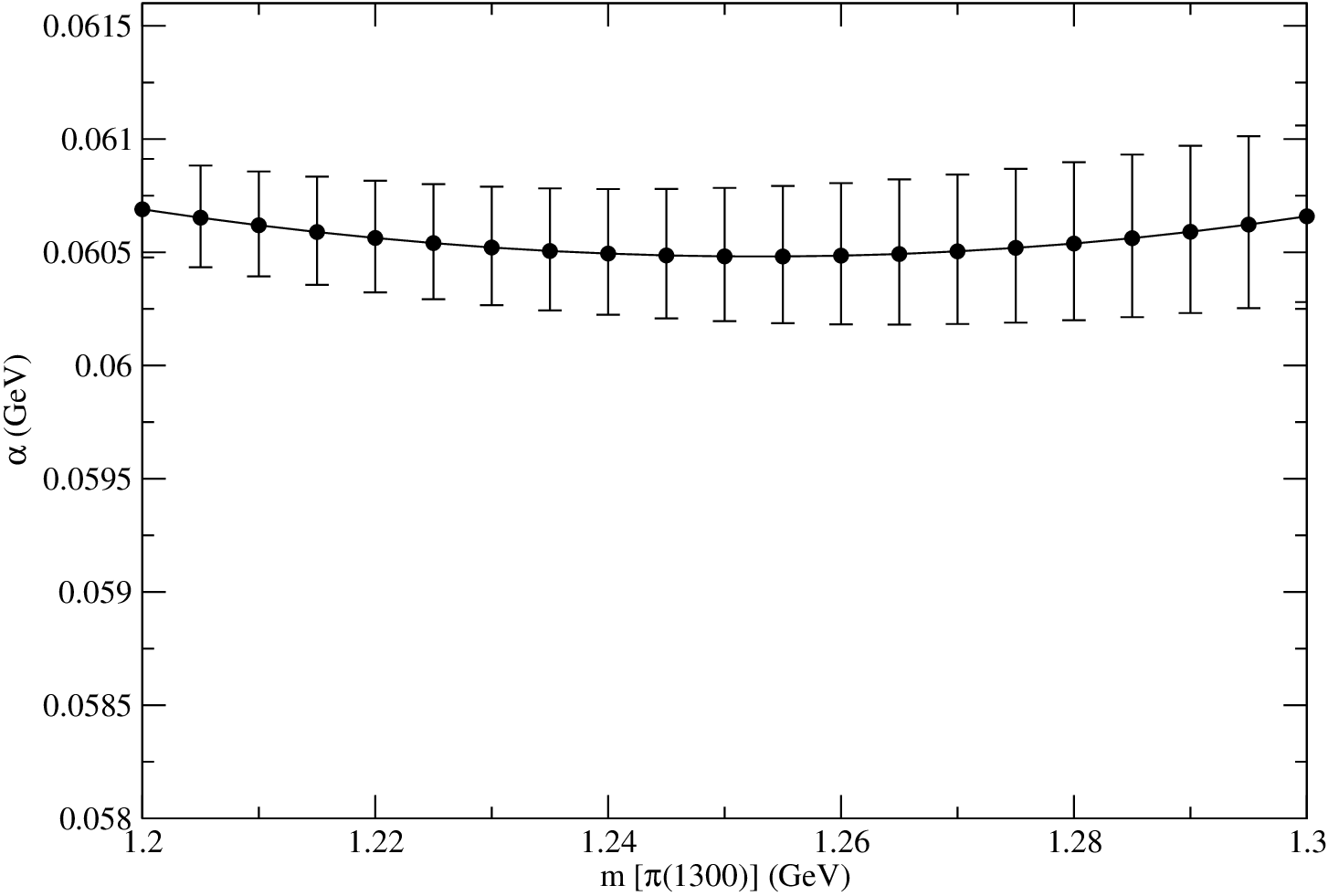}
\hskip 1cm
\epsfxsize = 7.5cm
\epsfbox{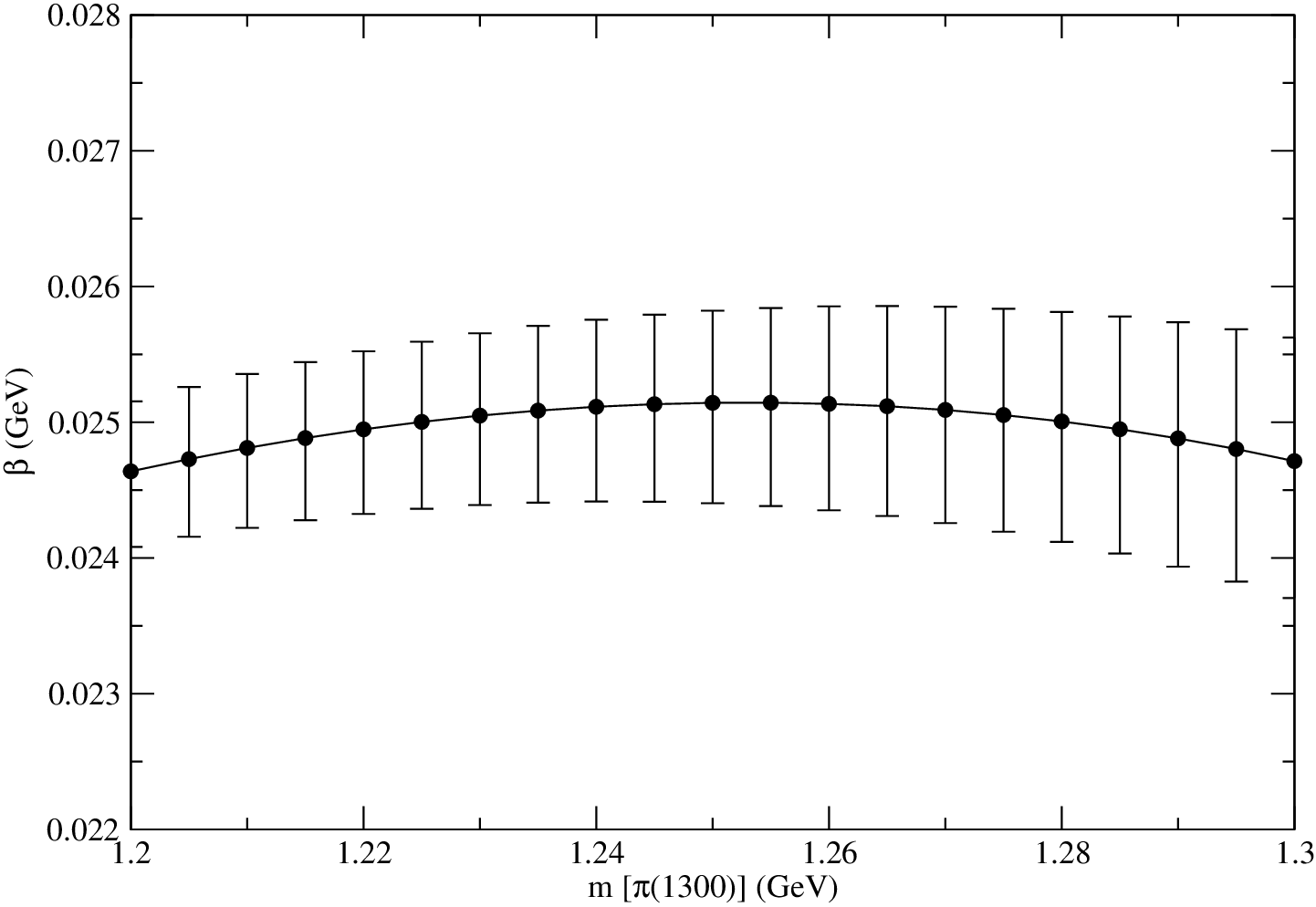}
\end{center}
\caption[]{%
Dependences of the two quark vacuum value $\alpha$ (left)
and the four quark vacuum value $\beta$ (right) on the choice of
$m[\pi(1300)]$. The error bars give the effect of the
uncertainty in the $a_0(1450)$ mass. }
\label{albe}
\end{figure}

     The four independent parameters which appear in the Lagrangian
($c_2$, $d_2$, $c_4^a$, $e_3^a$)
are shown, as functions of $m[\pi(1300)]$, in Fig. \ref{ind4}.
The vacuum expectation values $\alpha$ and $\beta$ of the two and four
quark scalar fields are similarly shown in Fig. \ref{albe}. It is seen
that $\beta$ and $\alpha$ are each insensitive to varying
$m[\pi(1300)]$ and their ratio is about 0.40.

\begin{figure}[t]
\begin{center}
\vskip 1cm
\epsfxsize = 7.5cm
\ \epsfbox{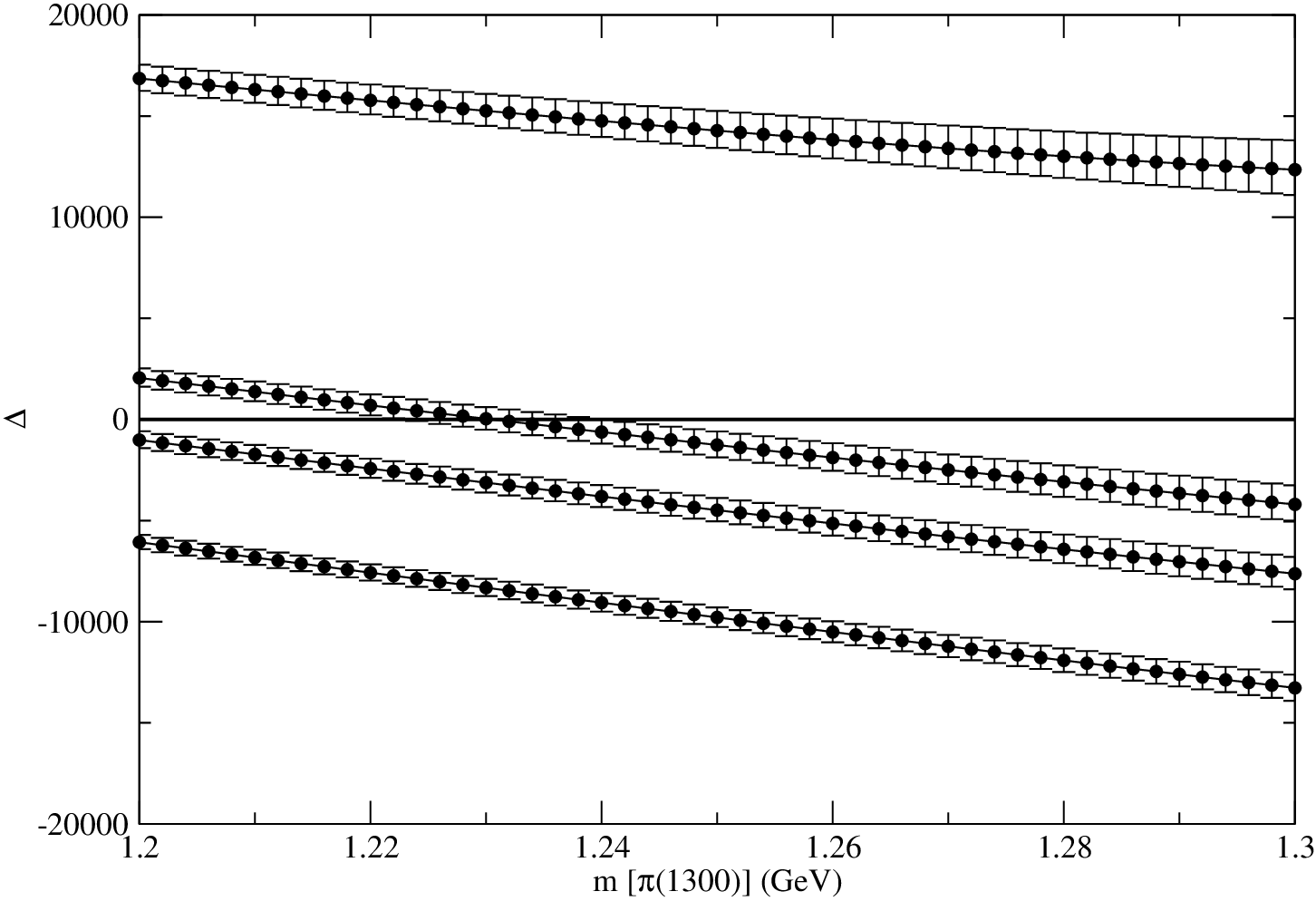}
\end{center}
\caption[]{%
The discriminant for Eq.(\ref{findgamma1})
vs. $m[\pi(1300)]$. The curves from bottom to top
respectively represent the choices for the heavier $0^-$ SU(3)
singlet to be
$\eta(1295)$, $\eta(1405)$, $\eta(1475)$ and $\eta(1760)$.
The error bars give the effect of the
uncertainty in the $a_0(1450)$ mass.
}
\label{disc}
\end{figure}

       To calculate the masses of the SU(3) singlet pseudoscalars
we must diagonalize Eq.(\ref{phizeromixing})
with the specific choices of parameters $y_0=2d_2$ and
$z_0=4e_3^a\alpha/d_2$ corresponding to the potential of
Eq.(\ref{SpecLag}). This enables us to fit in principle,
for any choice of $m[\pi(1300)]$, the two
parameters $c_3$ and $\gamma_1$ in terms
of the experimental masses of  $\eta$(958) and one of
the candidates $\eta(1295)$, $\eta(1405)$, $\eta(1475)$
and $\eta(1760)$. The specific formulas are given as
Eqs. (\ref{findgamma1}) and (\ref{findc3}) in Appendix B.
However, as mentioned above, the
positivity of the eigenvalues of
the matrix $(M_0^2)$ imposes additional constraints on the choice of
$m[\pi(1300)]$ in Eq.(\ref{scalarconscond}). This appears in solving
for $\gamma_1$ using the quadratic equation (\ref{findgamma1}) and
requiring its discriminant to be positive. In Fig. \ref{disc}, the
discriminants are shown as functions of $m[\pi(1300)]$ for each of the
four possible candidates for the heavier $0^-$ SU(3) singlet.
This clearly shows that the two lowest mass candidates have
negative discriminants and can be ruled out according to our criterion.
The perhaps most likely candidate $\eta(1475)$
[this case will be denoted scenario 1] has a positive
discriminant for $m[\pi(1300)]$ less than about 1.23 GeV. This leads to
the modified allowed ranges for the $0^+$ singlet states, shown in
parentheses in Eq.(\ref{scalsing}). There is no restriction on the
heaviest candidate, $\eta(1760)$ [this case will be denoted scenario 2].

\begin{figure}[h]
\begin{center}
\epsfxsize = 7.5cm
\ \epsfbox{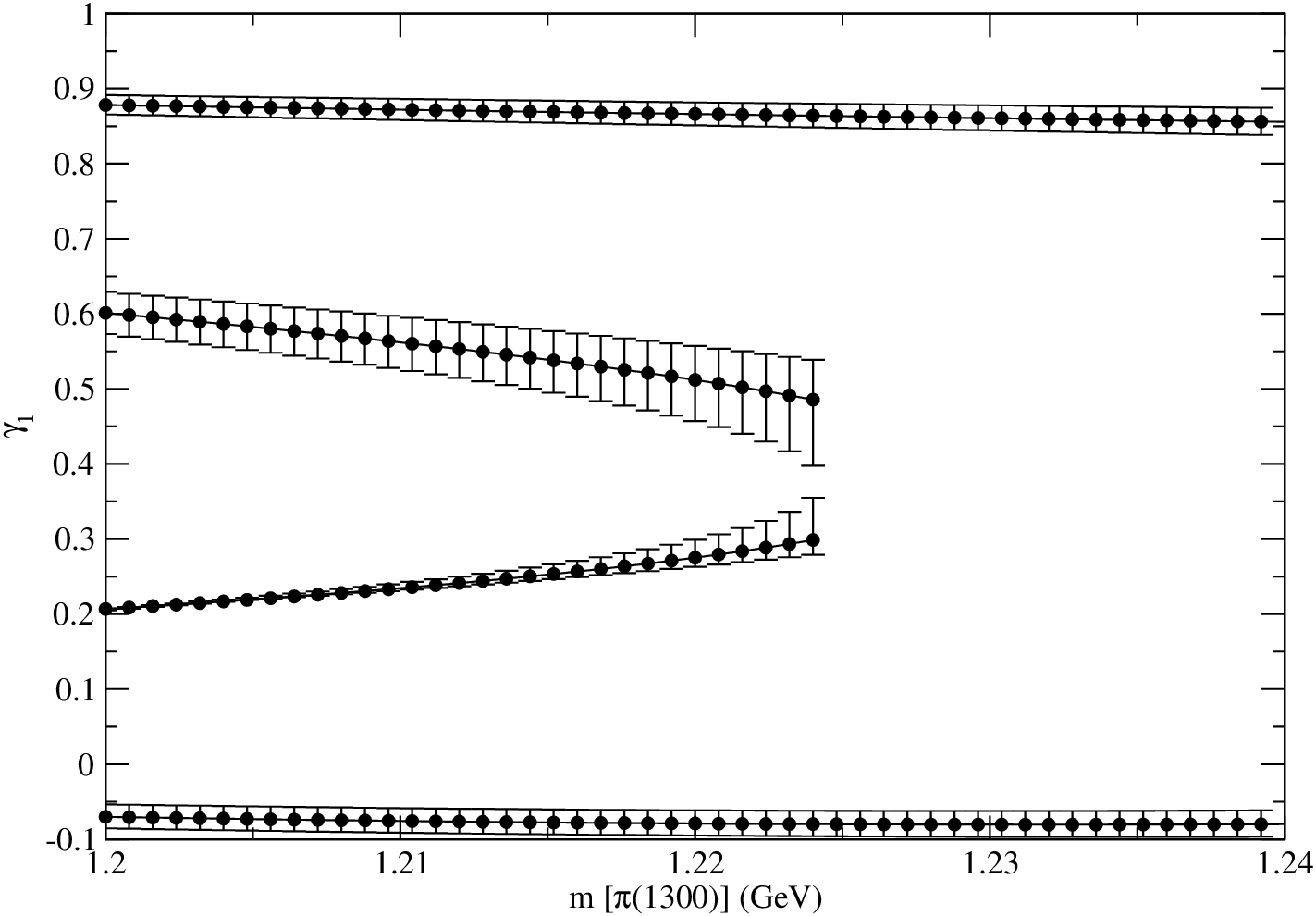}
\end{center}
\caption[]{%
$\gamma_1$ vs. $m[\pi(1300)]$.
The top and bottom curves correspond to choosing
the $\eta(1760)$ as the heavier $0^-$ SU(3) singlet while the middle two
curves correspond
to choosing the $\eta(1475)$ as the heavier $0^-$ SU(3) singlet.
Note that for each scenario, the two curves are associated with different
solutions of the quadratic equation \ref{findgamma1} for $\gamma_1$.
The error bars give the effect of the
uncertainty in the $a_0(1450)$ mass.
}
\label{gamma1plot}
\end{figure}

\begin{figure}[h]
\begin{center}
\vskip 1cm
\epsfxsize = 7.5cm
\ \epsfbox{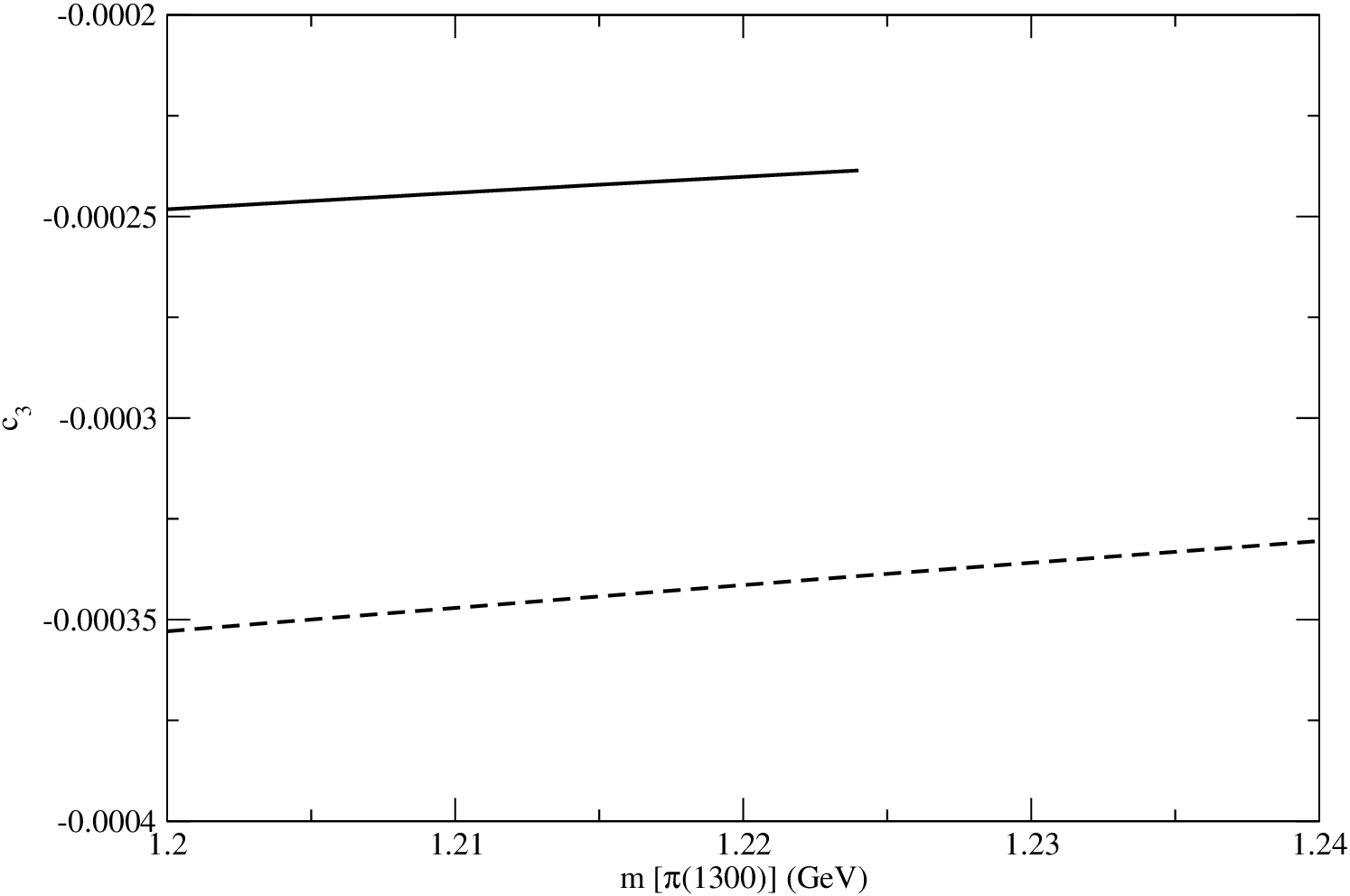}
\end{center}
\caption[]{%
$c_3$ in units of GeV$^4$ vs.  $m[\pi(1300)]$. The upper curve
corresponds to the scenario where the heavier $0^-$ SU(3) singlet is
identified
with the $\eta(1475)$ while the lower curve corresponds to $\eta(1760)$
as the heavier $0^-$ SU(3) singlet.
}
\label{c3plot}
\end{figure}

    Since Eq.(\ref{findgamma1}) is a quadratic equation for $\gamma_1$,
one expects that there may be two physical solutions for $\gamma_1$.
This turns out to be the case. In Fig.\ref{gamma1plot} we show plots of
$\gamma_1$
as a function of $m[\pi(1300)]$ for each of the scenarios mentioned
above. The quantity $c_3$ is given in Eq.(\ref{findc3}) and is seen
in Fig.\ref{c3plot} to be single valued in its dependence on
$m[\pi(1300)]$.

\begin{figure}[h]
\begin{center}
\epsfxsize = 7.5cm
\ \epsfbox{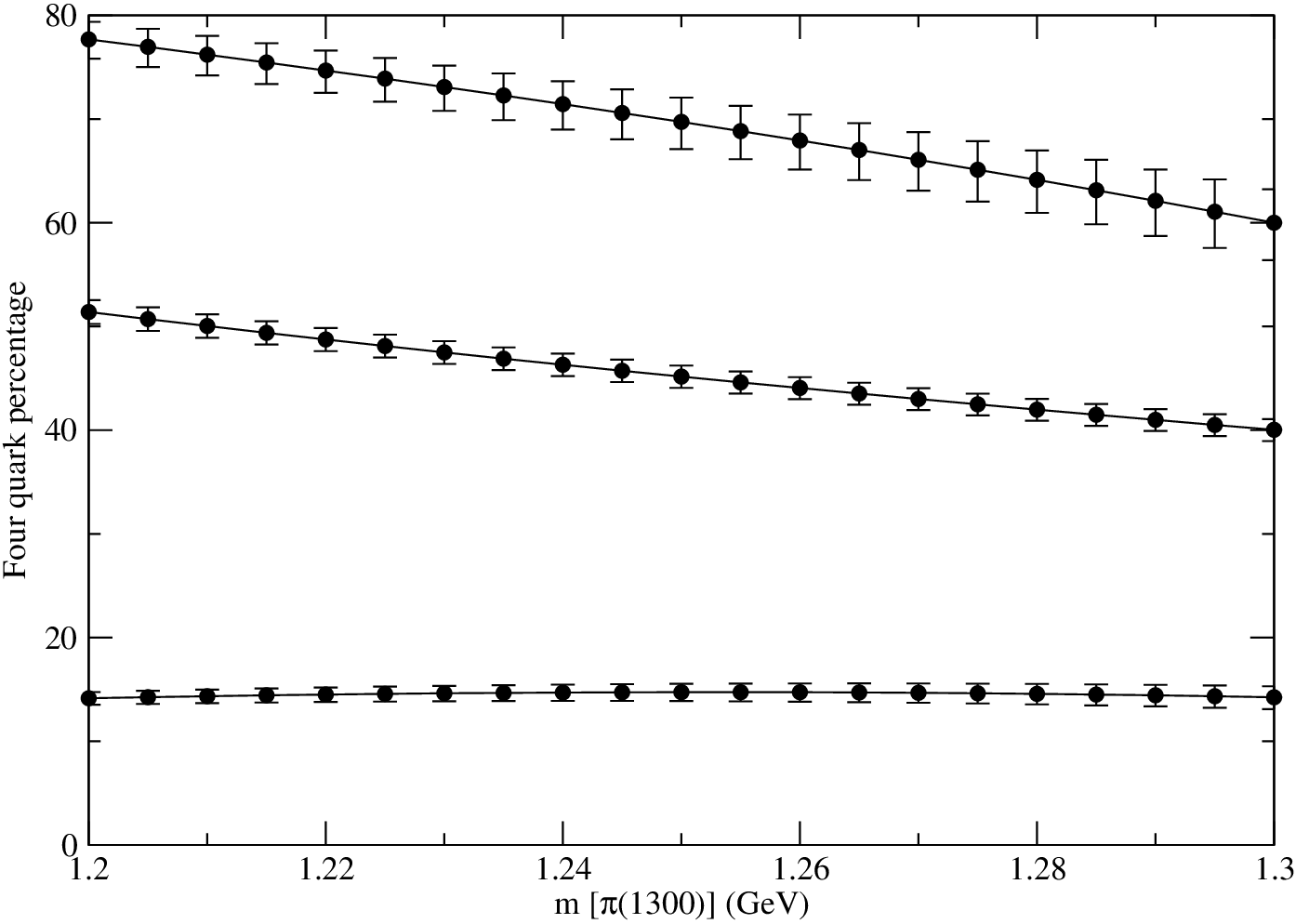}
\end{center}
\caption[]{%
Plot of the four quark percentages of various particles in the model
as functions of the undetermined input parameter, $m[\pi(1300)]$.
Starting from the bottom and going up, the curves respectively
show the four quark percentages of the pion, the $0^+$ singlet,
and the $a_0(980)$.
The error bars give the effect of the
uncertainty in the $a_0(1450)$ mass.
}
\label{4qp}
\end{figure}

    It is very interesting to see what the model has to say about the
four
quark percentages of the particles it describes. The percentages
for the pion, the lighter $0^+$ singlet and the $a_0(980)$ are
displayed in Fig.\ref{4qp} as functions of the precise value of the
input parameter $m[\pi(1300)]$. The pion four quark content (equal
to 100 ${\rm sin} ^2\theta_\pi$) is seen to be about 17
percent. Of course the
heavier pion would have about an 83 percent four quark content. On the
other hand, the octet scalar states present a reversed picture: the
$a_0(980)$ has a large four quark content while the $a_0(1450)$ has a
smaller four quark content. The very light and the rather heavy
$0^+$ singlets are about maximally mixed, having roughly equal
contributions from the 4 quark and 2 quark components.

\begin{figure}[h]
\begin{center}
\vskip 1cm
\epsfxsize = 7.5cm
\ \epsfbox{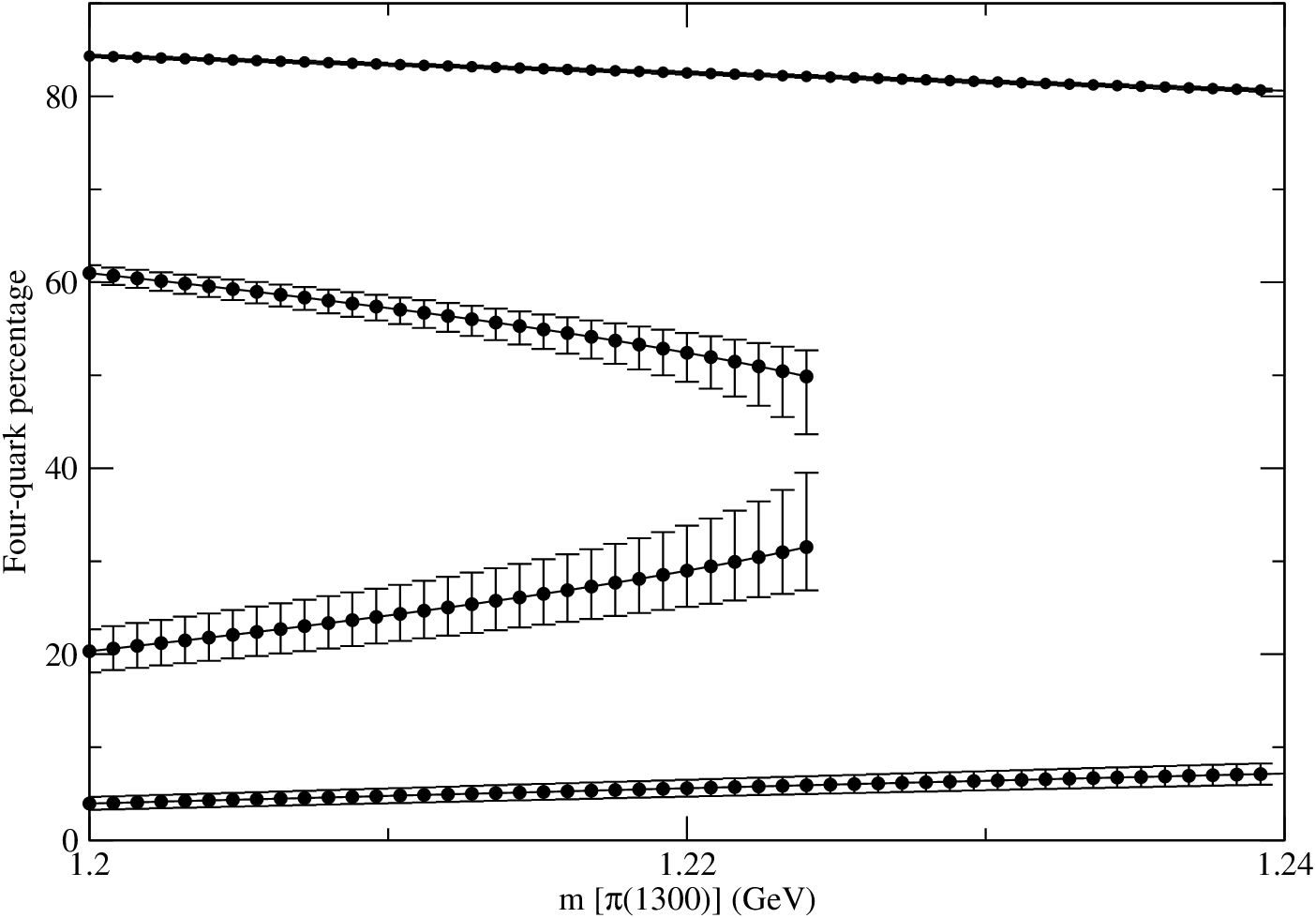}
\end{center}
\caption[]{%
Plot of the four quark percentages of
the $\eta(958)$
as functions of the undetermined input parameter, $m[\pi(1300)]$
for two scenarios. The top and bottom curves correspond to choosing
the $\eta(1760)$ as the heavier $0^-$ SU(3) singlet while the middle two
curves correspond
to choosing the $\eta(1475)$ as the heavier $0^-$ SU(3) singlet.
Note that for each scenario, the two curves are associated with different
solutions of the quadratic equation \ref{findgamma1} for $\gamma_1$.
The error bars give the effect of the
uncertainty in the $a_0(1450)$ mass.
}
\label{4qpeta}
\end{figure}

   In Fig.\ref{4qpeta} the four quark percentages of the $0^-$
SU(3) singlets are shown for both scenarios.
The perhaps more
plausible scenario takes
$\eta(1475)$ as the heavy
$0^-$ singlet state. In this case we see that for the solution with
smaller
$\gamma_1$, the four quark content of the
familiar $\eta(958)$ is about 25 percent while for the solution with
larger $\gamma_1$, the four quark content of $\eta(958)$ is about
55 percent. Thus the smaller $\gamma_1$ solution seems more plausible
physically. In the case where the $\eta(1760)$ is identified
as the heavier partner of the $\eta(958)$ the smaller $\gamma_1$
solution yields an $\eta(958)$ with a four quark content of about 7
percent while the larger $\gamma_1$ solution yields an $\eta(958)$
with a four quark content of about 82 percent.

    Values of all the model parameters as well as numerical values
of the mixing matrices, for a typical choice of $m[\pi(1300)]$, are
listed at the end of Appendix B.

\section{Three point vertices}
    The three point vertices are useful for calculating the widths
of the various mesons and also for the calculation of meson-meson
scattering. These can be calculated for a specific model, like the
one with the choice
of terms given in the previous section, by straightforward
differentiation. However, one may also obtain model independent
(in the sense of being independent of the choice of
invariant terms in $V_0$) information
about these from the generating equation. We shall do that
here, specializing to the scalar-pseudoscalar-pseudoscalar vertices
needed for pion pion scattering. These are obtained by
succesively differentiating the
two equations in Eq.(\ref{geneqs}) with respect to one
scalar field and one
pseudoscalar field. First we introduce the notations:
\begin{eqnarray}
r_1=\left\langle {{\partial^3 V_0}\over{\partial\phi_2^1\partial
\phi_1^2\partial S_0}} \right\rangle \hspace{1cm} q_1=\left\langle
{{\partial^3
V_0}\over{\partial\phi_2^1\partial \phi_1^2\partial S_8}} \right\rangle
\nonumber\\
r_2=\left\langle {{\partial^3 V_0}\over{\partial\phi_2^1\partial
\phi_1^2\partial S'_0}} \right\rangle \hspace{1cm} q_2=\left\langle
{{\partial^3
V_0}\over{\partial\phi_2^1\partial \phi_1^2\partial {S'}_8}} \right\rangle
\nonumber\\
r_3=\left\langle {{\partial^3 V_0}\over{\partial\phi_2^1\partial
{\phi'_1}^ 2\partial S_0}} \right\rangle \hspace{1cm} q_3=\left\langle
{{\partial^3 V_0}\over{\partial\phi_2^1\partial {\phi'}_1^2\partial
S_8}} \right\rangle
\nonumber\\
r_4=\left\langle {{\partial^3 V_0}\over{\partial\phi_2^1\partial
{\phi'}_1^2\partial {S'}_0}} \right\rangle \hspace{1cm} q_4=\left\langle
{{\partial^3 V_0}\over{\partial\phi_2^1\partial {\phi'}_1^2\partial
{S'}_8}} \right\rangle
\nonumber\\
r_5=\left\langle {{\partial^3 V_0}\over{\partial{\phi'}_2^1\partial
{\phi'}_1^2\partial S_0}} \right\rangle \hspace{1cm} q_5=\left\langle
{{\partial^3 V_0}\over{\partial{\phi'}_2^1\partial
{\phi'}_1^2\partial S_8}} \right\rangle
\nonumber\\
r_6=\left\langle {{\partial^3 V_0}\over{\partial{\phi'}_2^1\partial
{\phi'}_1^2\partial {S'}_0}} \right\rangle \hspace{1cm} q_6=\left\langle
{{\partial^3 V_0} \over{\partial{\phi'}_2^1\partial
{\phi'}_1^2\partial {S'}_8}} \right\rangle
\label{rq}
\end{eqnarray}
Note that $S_0$ was defined in Eq.(\ref{singlets}) while $S_8$, for
example, is
the isoscalar member of the SU(3) octet defined as:
\begin{equation}
S_8=\frac{1}{\sqrt{6}}(S_1^1+S_2^2-2S_3^3).
\label{s8}
\end{equation}
Now using the generating equations as just discussed, we obtain the
following relations connecting
the trilinear coupling constants $r_i$ with corresponding
mass squared matrices for the $S_0$-$S'_0$ and the
$\pi$-$\pi'$ systems.

\begin{eqnarray}
\alpha r_1+\beta r_3&=&\frac{1}{ \sqrt {3}}\left\langle{{\partial^2
V_0}\over{\partial
S_0^2}}\right\rangle-\frac{1}{\sqrt{3}}\left\langle{{\partial^2
V_0 }\over
{\partial {\phi}_2^1\partial{\phi}_1^2}}\right\rangle
\nonumber\\
\alpha r_2+\beta r_4&=&\frac{1}{\sqrt{3}}\left\langle{{\partial^2
V_0}\over{\partial S_0\partial
{S'}_0}}\right\rangle-\frac{1}{\sqrt{3}}\left\langle{{\partial^2 V_0
}\over
{\partial
{\phi}_2^1\partial{\phi'}_1^2}}\right\rangle
\nonumber\\
\alpha r_3+\beta r_5&=&\frac{1}{\sqrt{3}}\left\langle{{\partial^2
V_0}\over{\partial S_0\partial
{S'}_0}}\right\rangle-\frac{1}{\sqrt{3}}\left\langle{{\partial^2 V_0
}\over
{\partial
{
\phi}_2^1\partial{\phi'}_1^2}}\right\rangle
\nonumber\\
\alpha r_4+\beta r_6&=&\frac{1}{\sqrt{3}}\left\langle{{\partial^2
V_0}\over{\partial
{S'}_0^2}}\right\rangle-\frac{1}{\sqrt{3}}\left\langle{{\partial^2
V_0}\over {\partial {\phi'}_2^1\partial{\phi'}_1^2}}\right\rangle
\label{tritwo}
\end{eqnarray}
Similar equations are obtained for the $q_i$ trilinear couplings and
the mass squared matrices for the $S_8$-$S'_8$ systems:
\begin{eqnarray}
\alpha q_1+\beta q_3&=&\frac{1}{ \sqrt {6}}\left\langle{{\partial^2
V_0}\over{\partial
S_8^2}}\right\rangle-\frac{1}{\sqrt{6}}\left\langle{{\partial^2
V_0 }\over
{\partial \phi_2^1\partial\phi_1^2}}\right\rangle
\nonumber \\
\alpha q_2+\beta q_4&=&\frac{1}{\sqrt{6}}\left\langle{{\partial^2
V_0}\over{\partial S_8\partial
{S'}_8}}\right\rangle-\frac{1}{\sqrt{6}}\left\langle{{\partial^2 V_0
}\over
{\partial
\phi_2^1\partial{\phi'}_1^2}}\right\rangle
\nonumber \\
\alpha q_3+\beta q_5&=&\frac{1}{\sqrt{6}}\left\langle{{\partial^2
V_0}\over{\partial S_8\partial
{S'}_8}}\right\rangle-\frac{1}{\sqrt{6}}\left\langle{{\partial^2 V_0
}\over
{\partial
\phi_2^1\partial{\phi'}_1^2}}\right\rangle
\nonumber \\
\alpha q_4+\beta q_6&=&\frac{1}{\sqrt{6}}\left\langle{{\partial^2
V_0}\over{\partial
{S'}_8^2}}\right\rangle-\frac{1}{\sqrt{6}}\left\langle{{\partial^2
V_0
}\over {\partial {\phi'}_2^1\partial{\phi'}_1^2}}\right\rangle
\label{tritwo2}
\end{eqnarray}

Eq.(\ref{tritwo}) and Eq.(\ref{tritwo2}) relate
eight different linear combinations of the three point
vertices to two
point vertices for the fields of pure $q{\bar q}$ and pure $qq{\bar
q}{\bar q}$ types. Since there are twelve a priori unknown
three point vertices according to Eq.(\ref{rq}), it is clear that there
is, in general, not enough information available to determine all the
three point vertices in terms of the two point ones.
However, we will see that the available relations are sufficient to
prove the desired low energy theorem.
To relate the quantities in
Eqs.(\ref{tritwo}) and (\ref{tritwo2})
to quantities pertaining to mass
eigenstates we introduce an index notation to distinguish unprimed from
primed fields; for example:
\begin{equation}
\phi_1^2 = ({\phi}_1^2)_1, \hspace{2cm} {\phi'}_1^2=({\phi}_1^2)_2.
\label{notation}
\end{equation}
With this notation, which we apply to all fields of the model, the
coupling constant of the Goldstone boson
pions to the mass diagonal SU(3) singlet scalars may be compactly written
as:
\begin{equation}
g_{0D}=
\left\langle
{
  {{\partial^3V}}
       \over
  {\partial{\pi}^+\partial{\pi}^- \partial(S_{0p})_D  }
}
\right\rangle =\sum_{A,B,C}(R_\pi)_{A1}(R_\pi)_{B1} (L_0)_{CD}
\left\langle{{\partial^3V}\over{\partial({\phi}_1^2)_A\partial({\phi}_2^1)_B
\partial(S_0)_C}}\right\rangle.
\label{chainrule}
\end{equation}
The transformation matrix elements, $(R_\pi)_{AB}$ and $(L_0)_{AB}$
may be read
from Eq.(\ref{mixingangle}) and Eq.(\ref{szeromixingangle}). Note that
the capital Latin subscripts take on the values 1 and 2
as in Eq.(\ref{notation}) above.
There is a similar equation involving the
$S_8$-$S'_8$ scalars which yields the physical coupling constant of two
Goldstone pions with $S_8$, $g_{8D}$:
\begin{equation}
g_{8D}=
\left\langle
{
  {{\partial^3V}}
       \over
  {\partial{\pi}^+\partial{\pi}^- \partial(S_{8p})_D  }
}
\right\rangle =\sum_{A,B,C}(R_\pi)_{A1}(R_\pi)_{B1}(L_a)_{CD}
\left\langle{{\partial^3V}\over{\partial({\phi}_1^2)_A\partial({\phi}_2^1)_B
\partial(S_8)_C}}\right\rangle.
\label{chainrule8}
\end{equation}
Here, $L$ is the transformation matrix defined in Eq.(\ref{amixingangle}).
   Using the compact form of Eq.(\ref{chainrule}), one may compactly
express the comparison of Eq.(\ref{tritwo}) with Eq.(\ref{rq}) as:
\begin{equation}
\frac{\sqrt{3}F_\pi}{2}\sum_{B}(R_\pi^{-1})_{1B}
\left\langle{{\partial^3V_0}\over{\partial({\phi}_1^2)_A\partial({\phi}_2^1)_B
\partial(S_0)_H}}\right\rangle
= (X_0^2)_{AH}-(M_\pi^2)_{AH}.
\label{3to2}
\end{equation}
$(M_\pi^2)$ is given in Eq.(\ref{mpipiprime}) and
$(X_0^2)$
is the model independent version of Eq.(\ref{s0mass}).
Similarly,
\begin{equation}
\frac{\sqrt{6}F_\pi}{2}\sum_{B}(R_\pi^{-1})_{1B}
\left\langle{{\partial^3V_0}\over{\partial({\phi}_1^2)_A\partial({\phi}_2^1)_B
\partial(S_8)_H}}\right\rangle
= (X_a^2)_{AH}-(M_\pi^2)_{AH}.
\label{second3to2}
\end{equation}
Here $(X_a^2)$
is the model independent version of Eq.(\ref{s8mass}).
Note that according to our conventions the nondiagonal
and diagonal (hatted) squared mass matrices are related
as:
\begin{eqnarray}
\sum_{B,C}(R_{\pi}^{-1})_{AB}(M_{\pi}^2)_{BC}(R_{\pi})_{CD}&=&({\hat
M}_{\pi}^2)_{AD},\hskip 0.5cm
\sum_{B,C}(R_0^{-1})_{AB}(M_0^2)_{BC}(R_0)_{CD}=({\hat
M}_0^2)_{AD},
\nonumber \\
\sum_{B,C}(L_a^{-1})_{AB}(X_a^2)_{BC}(L_a)_{CD}&=&({\hat
X}_a^2)_{AD}, \hskip 0.5cm
\sum_{B,C}(L_0^{-1})_{AB}(X_0^2)_{BC}(L_0)_{CD}=({\hat
X}_0^2)_{AD},
\label{simtransf}
\end{eqnarray}

\section{Low energy pion scattering}
    There are two reasons for next discussing the pi-pi
scattering in this model. First, since the iso-singlet scalar
resonances above are being considered at tree level, one expects,
as can be seen in the single $M$ model also discussed in \cite{BFMNS01}
and at the two flavor level in \cite{AS94},
that unitarity corrections for the scattering amplitudes will
alter their masses and widths. Second, since the pion looks
unconventional in this model (having a non-neglegible four quark
component) one might worry that the fairly precise ``current algebra"
formula for the near to threshold scattering amplitude might acquire
unacceptably large corrections.

    Of course, for computing the near threshold pion pion scattering,
it is well known that the use of a nonlinear sigma model is more
convenient. However, we are also interested in unitarizing the model in
the resonance region where the nonlinear model, which can be obtained by
integrating out the resonances, is clearly not applicable.

    The invariant pion pion scattering amplitude for
$\pi_i(p_1)$+$\pi_j(p_2)$$\rightarrow$ $\pi_k(p_3)$+$\pi_l(p_4)$ is
decomposed as:
\begin{equation}
\delta_{ij}\delta_{kl}A(s,t,u) +
\delta_{ik} \delta_{jl}A(t,s,u) +
\delta_{il} \delta_{jk}A(u,t,s),
\label{invamp}
\end{equation}
where $s$, $t$ and $u$ are the usual Mandelstam variables.
Note that the phase of the above amplitude simply corresponds to taking
the matrix element of the Lagrangian density for a four pion contact
interaction. The $I=0$, $I=1$ and $I=2$ amplitudes correspond to the
projections:
\begin{eqnarray}
T^0(s,t,u)&=&3A(s,t,u)+A(t,s,u)+A(u,t,s),
\nonumber \\
T^1(s,t,u)&=&A(t,s,u)-A(u,t,s),
\nonumber \\
T^2(s,t,u)&=&A(t,s,u)+A(u,t,s).
\label{isoprojections}
\end{eqnarray}

It is straightforward to calculate $A(s,t,u)$ using the three
point vertices for two massless pions coupling to a physical scalar
(See Eqs. (\ref{chainrule}) and (\ref{chainrule8})) as well as the four
point coupling constant, $g$
for four massless pions:
\begin{equation}
g= \left\langle {
  {\partial^4V_0}
\over{\partial{\pi}^+\partial{\pi}^- \partial{\pi}^+
\partial{\pi}^-} } \right\rangle
\label{4ptcoupling}
\end{equation}
The result is simply:
\begin{equation}
A(s,t,u)=-\frac{g}{2} +\sum_D
\left(\frac{g_{8D}^2}{({\hat X}_a^2)_{DD}-s}
+\frac{g_{0D}^2}{({\hat X}_0^2)_{DD}-s}\right).
\label{wholeamp}
\end{equation}
Note that the sum goes over the two SU(3) singlet scalars as well as the
two iso-singlet scalars belonging to SU(3) octets.
We are presently interested in the threshold region (near $s$=0 for
massless pions)
so we expand this formula to first order in s:
\begin{equation}
A(s,t,u)\approx-\frac{g}{2} +\left(\frac{g_{8D}^2}{({\hat X}_a^2)_{DD}}
+\frac{g_{0D}^2}{({\hat X}_0^2)_{DD}}\right)+
s\left(\frac{g_{8D}^2}{[({\hat X}_a^2)_{DD}]^2}
+\frac{g_{0D}^2}{[({\hat X}_0^2)_{DD}]^2}\right).
\label{expandedamp}
\end{equation}
In this equation the summation over $D$ has not been explicitly written
and the summation over repeated indices is to be assumed; note that the
quantity $({\hat X}_a^2)_{DD}$, for example, is a single number indexed
by $D$.
Observe that the four point vertex does not contribute to the terms linear
in $s$. Let us then evaluate the $s$ term first. Begin by substituting
Eq.(\ref{3to2}) into Eq.(\ref{chainrule}) and noticing that the term
$(M_\pi^2)_{AH}$ makes zero contribution since that piece can be
manipulated, using Eq.(\ref{simtransf}), to be proportional to the zero
masses of the
physical Goldstone bosons. The physical trilinear coupling constant is
next obtained as $g_{0D}=\frac{2}{\sqrt{3}F_\pi}
(R_\pi)_{A1}(X_0^2)_{AH}(L_0)_{HD}$.
Then the quantity appearing in
Eq.(\ref{expandedamp}) can be evaluated as:

\begin {eqnarray}
&&\frac{g_{0D}^2}{[({\hat
X}_0^2)_{DD}]^2}=\frac{4}{3F_\pi^2}(R_\pi)_{A1}(X_0^2)_{AH}(L_0)_{HD}
\frac{1}{[({\hat X}_0^2)_{DD}]^2}(R_{\pi})_{C1}(X_0^2)_{CK}(L_0)_{KD}
\nonumber \\
&&=\frac{4}{3F_{\pi}^2}(R_{\pi}^{-1})_{1G}(L_0)_{GE}(L_0^{-1})_{EA}(X_0^2)_{AH}
(L
_0)_{HD}\frac{1}{[({\hat X}_0^2)_{DD}]^2}
(L_0^{-1})_{DK}(X_0^2)_{KC}(L_0)_{CF}(L_0^{-1})_{FJ}(R_{\pi})_{J1}
\nonumber\\
&&=\frac{4}{3F_{\pi}^2}(R_{\pi}^{-1})_{1G}(L_0)_{GE}({\hat
X}_0^2)_{ED}\frac{1}{[({\hat X}_0^2)_{DD}]^2}({\hat X}_0^2)_{D
F}(L_0^{-1})_{FJ}(R_{\pi})_{J1}=
\frac{4}{3F_{\pi}^2}.
\label{g2m40}
\end{eqnarray}
Similarly,
\begin{equation}
\frac{g_{8D}^2}{[({\hat X}_a^2)_{DD}]^2}=\frac{4}{6F_{\pi}^2}
\label{g2m48}
\end{equation}

The $s$ dependent part of the scattering amplitude near threshold
finally takes the simple form:
\begin{equation}
A(s,t,u)= \frac{2s}{F_\pi^2}.
\label{CAresult}
\end{equation}

This may be recognized as the usual current algebra
formula \cite{W} in the case where the pion mass is set to zero.
We will complete its derivation in the next section, where it
will be shown that the s independent terms in
Eq.(\ref{expandedamp}) cancel each other. It should be
remarked that the present derivation holds for any choice of
chiral invariant terms in $V_0$, not necessarily just for
the leading terms in Eq.(\ref{SpecLag}).

     Of course, the current algebra result is just the first
term in an expansion in powers of $s$. The higher terms will have
the structure of a geometric series:
\begin{equation}
A(s,t,u)=s\left[\frac{2}{F_\pi^2}+
s\sum_i\frac{g_i^2}{m_i^6}+s^2\sum_i\frac{g_i^2}{m_i^8}+\cdots
\right],
\label{Aexpanded}
\end{equation}
wherein we have amalgmated all four scalars as the $m_i$ and
their corresponding coupling constants to two pions
as the $g_i$. It may be noted
that the entire amplitude is proportional to s.
The zero of the amplitude at
$s=0$ is referred to as the Adler zero. Notice also
that the higher terms involve the scalar masses and hence
will vanish as the $m_i\rightarrow\infty$. In the case of the
linear-in-s current algebra term, the non zero result arose
because the $g_i$'s increase as $m_i^2$. Taking the
scalar masses to infinity is the same as integrating them
out of the Lagrangian which results, as pointed out
in the original paper \cite{gl} by
Gell-Mann and Levy, in a nonlinear sigma model. The
magic cancellations in that case are very easy to see.
Clearly they are more intricate in the present case.

    From the starting equation (\ref{wholeamp}) it is seen
that the radius of convergence of the series
in $s$ is equal to the
squared mass of the lightest scalar meson. To go beyond this
point, in principle one should calculate all loop diagrams.
A simple approximation is to identify the partial wave
corresponding to the tree term with the K matrix amplitude.
This gives results
for amplitudes spanning a considerable range in $s$
in reasonable agreement with present experimental
indications. This was carried out for the SU(2) single M
linear sigma model in \cite{AS94} and for the SU(3) single M
linear sigma model in \cite{BFMNS01}.

\section{Four point vertices}
We start by establishing the notations for the quadrilinear
coupling constants involving the prediagonal fields:
\begin{eqnarray}
p_1&=&\left\langle {{\partial^4 V_0}
\over{\partial\phi_2^1\partial\phi_1^2\partial\phi_2^1\partial\phi_1^2}}
\right\rangle
\nonumber\\
p_2&=&\left\langle {{\partial^4 V_0}
\over{\partial{\phi'}_2^1\partial\phi_1^2\partial\phi_2^1\partial\phi_1^2}}
\right\rangle
\nonumber\\
p_{31}&=&\left\langle {{\partial^4 V_0}
\over{\partial{\phi'}_2^1\partial{\phi'}_1^2\partial\phi_2^1\partial\phi_1^2}}
\right\rangle
\nonumber\\
p_{32}&=&\left\langle {{\partial^4 V_0}
\over{\partial{\phi'}_2^1\partial\phi_1^2\partial{\phi'}_2^1\partial\phi_1^2}}
\right\rangle
\nonumber\\
p_4&=&\left\langle {{\partial^4 V_0}
\over{\partial{\phi'}_2^1\partial{\phi'}_1^2\partial{\phi'}_2^1\partial\phi_1^2
}}\right\rangle
\nonumber\\
p_5&=&\left\langle {{\partial^4 V_0}
\over{\partial{\phi'}_2^1\partial{\phi'}_1^2\partial{\phi'}_2^1\partial
{\phi'}_1^2}} \right\rangle
\label{p}
\end{eqnarray}
We find the following equations relating these quadrilinear
coupling constants to
the trilinear coupling constants in Eq.(\ref{rq}) by differentiating the
second generating equation in (\ref{geneqs}) three times with respect to
pseudoscalar fields :
\begin{eqnarray}
p_1&=&
\frac{\beta^2}{\alpha^2}p_{31}-\frac{\beta}{\sqrt{2}\alpha^2}
\left(\frac{1}{
\sqrt{3}}q_2+\frac{2}{\sqrt{6}}r_2
+\frac{1}{\sqrt{3}}q_3+\frac{2}{\sqrt{6}}r_3\right)
+\frac{\sqrt{2}}{\alpha}
\left( \frac{1}{\sqrt{3}}q_1+\frac{2}{\sqrt{6}}r_1\right)
\nonumber\\
p_2&=&-\frac{\beta}{\alpha}p_{31}+\frac{1}{\sqrt{2}\alpha}
\left( \frac{1}{\sqrt{3}}q
_2+\frac{2}{\sqrt{6}}r_2
+\frac{1}{\sqrt{3}}q_3+\frac{2}{\sqrt{6}}r_3\right)
\nonumber\\
p_{32}&=&p_{31}+\frac{1}{\beta\sqrt{2}}
\left( \frac{1}{\sqrt{3}}(q_3-q_2)+\frac{2}
{\sqrt{6}}(r_3-r_2)\right)
\nonumber\\
p_4&=&-\frac{\alpha}{\beta}p_{31}+\frac{1}{\sqrt{2}\beta}
\left(\frac{1}{\sqrt{3}}q_
4+\frac{2}{\sqrt{6}}r_4
+\frac{1}{\sqrt{3}}q_5+\frac{2}{\sqrt{6}}r_5\right)
\nonumber\\
p_5&=&\frac{\alpha^2}{\beta^2}p_{31}-\frac{\alpha}{\sqrt{2}\beta^2}
\left(
\frac{1}{\sqrt{3}}q_2+\frac{2}{\sqrt{6}}r_2
+\frac{1}{\sqrt{3}}q_3+\frac{2}{\sqrt{6}}r_3\right)
+\frac{\sqrt{2}}{\beta}
\left( \frac{1}{\sqrt{3}}q_6+\frac{2}{\sqrt{6}}r_6\right)
\label{pqr}
\end{eqnarray}

Notice that the above equations were obtained by expressing five out of
the six quantities in Eq.(\ref{p}) in terms of trilinear coupling
constants
as well as the sixth quadrilinear, $p_{31}$. This shows that all the
quadrilinear coupling constants cannot be obtained in terms of the
trilinear ones. Nevertheless,
as we will now see, the physical quadrilinear coupling
constant, $g$ can be completely expressed in terms of the bilinear
coupling constants. Using
the definition in
Eq.(\ref{4ptcoupling}), we express the physical four point coupling
constant in terms of the bare four point coupling constants as,
\begin{equation}
g=(R_\pi)_{A1}(R_\pi)_{B1}(R_\pi)_{C1}(R_\pi)_{D1}
\left\langle\frac{{\partial}^4V_0}{\partial (\phi^2_1)_A
\partial (\phi^1_2)_B\partial (\phi^2_1)_C\partial (\phi^1_2)_D }
\right\rangle,
\label{expandg}
\end{equation}
which may be explicitly written as,
\begin{equation}
g={\rm \cos}^4{\theta_\pi}p_1- 4{\rm \cos}^3{\theta_\pi}{\rm
sin}{\theta_\pi}p_2 +{\rm \cos}^2{\theta_\pi}{\rm
sin}^2{\theta_\pi}(4 p_{31}+2p_{32})- 4{\rm \cos}{\theta_\pi}{\rm
sin}^3{\theta_\pi}p_4
+ {\rm sin}^4
{\theta_\pi}p_5.
\label{g}
\end{equation}
Substituting Eq.(\ref{pqr}) into Eq.(\ref{g})
and then using Eqs.(\ref{tritwo}) and (\ref{tritwo2}) gives the
formula for the quadrilinear coupling constant:
\begin{eqnarray}
g&=&\frac{1}{(\alpha^2+\beta^2)^2}
\left[  \frac{2}{3}(\alpha^2(X_0^2)_{11}
+2\alpha\beta
(X_0^2)_{12}+\beta^2 (X_0^2)_{22}) \right.
\nonumber\\
&+&
\left.
\frac{1}{3}(\alpha^2(X_a^2)_{11}+2\alpha\beta
\left(X_a^2)_{12}+\beta^2(X_a^2)_{22}\right)\right] \label{finalg}
\end{eqnarray}
Noting $\alpha=(R^{-1}_\pi)_{11}F_\pi/2$ and
$\beta=(R^{-1}_\pi)_{12}F_\pi/2$ we rewrite Eq.(\ref{finalg})
as,
\begin{equation}
g=\frac{8}{F_{\pi}^2}\left( \frac{1}{3}
(R_{\pi}^{-1})_{1D}(
X_0^2)_{DJ}(R_{\pi})_{J1}+\frac{1}{6}(R_{\pi}^{-1})_{1D}(
X_0^2)_{DJ}(R_{\pi})_{J1}\right).
\label{gfinal}
\end{equation}
In order to verify the cancellation of the s independent terms in
Eq.(\ref{expandedamp})
we should subtract half of Eq.(\ref{gfinal}) from the sum of the following
two expressions:
\begin{eqnarray}
\frac{g_{0D}^2}{({\hat X}_0^2)_{DD}}&=&
\frac{4}{3F_{\pi}^2}(R_\pi)_{A1}(X_0^2)_{AH}(L_0)_{HG}
\frac{1}{({\hat X}_0^2)_{GG}}(R_{\pi})_{C1}(X_0^2)_{CK}(L_0)_{KG}
\nonumber\\
&=&\frac{4}{3F_{\pi}^2}(R_{\pi}^{-1})_{1D}(L_0)_{DE}(L_0^{-1})_{EA}(X_0^2)_{AH}
(L_0)_{HG}\frac{1}{({\hat X}_0^2)_{GG}}
(L_0^{-1})_{GK}(X_0^2)_{KC}(L_0)_{CF}(L_0^{-1})_{FJ}(R_{\pi})_{J1}
\nonumber\\
&=&\frac{4}{3F_{\pi}^2}(R_{\pi}^{-1})_{1D}(L_0)_{DE}({\hat
X}_0^2)_{EG} \frac{1}{({\hat X}_0^2)_{GG}} ({\hat
X}_0^2)_{GF}(L_0^{-1})_{FJ}(R_{\pi})_{J1}=
\frac{4}{3F_{\pi}^2}(R_{\pi}^{-1})_{1D}(
X_0^2)_{DJ}(R_{\pi})_{J1} \label{g2m20}
\end{eqnarray}

\begin{eqnarray}
\frac{g_{8D}^2}{({\hat X}_a^2)_{DD}}&=&
\frac{4}{6F_{\pi}^2}(R_\pi)_{A1}(X_a^2)_{AH}(L_0)_{HG}
\frac{1}{({\hat X}_a^2)_{GG}}(R_{\pi})_{C1}(X_a^2)_{CK}(L_0)_{KG}
\nonumber\\
&=&\frac{4}{6F_{\pi}^2}(R_{\pi}^{-1})_{1D}(L_0)_{DE}(L_0^{-1})_{EA}(X_a^2)_{AH}
(L_0)_{HG}\frac{1}{({\hat X}_a^2)_{GG}}
(L_0^{-1})_{GK}(X_a^2)_{KC}(L_0)_{CF}(L_0^{-1})_{FJ}(R_{\pi})_{J1}
\nonumber\\
&=&\frac{4}{6F_{\pi}^2}(R_{\pi}^{-1})_{1D}(L_0)_{DE}({\hat
X}_a^2)_{EG} \frac{1}{({\hat X}_a^2)_{GG}} ({\hat
X}_a^2)_{GF}(L_0^{-1})_{FJ}(R_{\pi})_{J1}=
\frac{4}{6F_{\pi}^2}(R_{\pi}^{-1})_{1D}(
X_a^2)_{DJ}(R_{\pi})_{J1} \label{g2m28}
\end{eqnarray}
It has thus been shown that the simple formula Eq.(\ref{CAresult})
holds
near threshold in the case of massless pions for an
arbitrary potential, $V_0$.

\section{Summary and discussion}

    We have given a detailed treatment of a systematic
approach to the
study of a linear sigma model containing one chiral nonet
transforming under U(1)$_A$ as a
quark-antiquark composite and
another chiral nonet transforming as a
diquark-anti diquark composite
(or, equivalently from a symmetry point of view, as
a two meson
molecule). Some highlights of this work
have been presented elsewhere \cite{FJS06}.
The model provides an intuitive explanation of a
current puzzle in low energy QCD: Recent work has
suggested the existence of a lighter than 1 GeV
nonet of scalar mesons which  behave like four quark
composites. On the other hand, the validity of a
spontaneously broken chiral symmetric description
would suggest that these states be (perhaps somewhat
distorted) chiral partners of the light
pseudoscalar mesons which are two quark composites.
The model solves the problem by starting with the
two chiral nonets mentioned and allowing them to mix
with each other. Working with the SU(3)
invariant version of the model it
is seen that the four experimental inputs given in
Eq.(\ref{inputs1}) (note that the lighter $0^-$ nonet
automatically has zero mass in the limit
in which we are working)
enforce a mixing whereby the light scalars
have a large four quark content while the light
pseudoscalars have a large
two quark content. In addition, one light isosinglet
scalar is exceptionally light (see Eq.(\ref{scalsing})).

    Of the four experimental inputs just mentioned, there
is a large uncertainty associated only with the mass of the
``heavy" pion, the $\pi(1300)$. It turns out that there
is in fact some sensitivity to the precise
choice of $m[\pi(1300)]$
so that this quantity is really being considered
as a free parameter within the range of the quoted
rather large experimental error. Thus the model parameters
and predictions calculated in
section IV are all displayed as functions of
$m[\pi(1300)]$. The effect of the not so large
allowed variations in the mass of the $a_0(1450)$ are
shown as error bars in these plots.

   In our treatment there are two parameters,
associated with the masses
and mixings of the SU(3) singlet pseudoscalars,
which describe the $U(1)_A$ anomaly in the
effective Lagrangian. These parameters do not affect
properties of the other particles and may be traded
for the masses of the $\eta(958)$ and one of the heavier
candidates $\eta(1295)$, $\eta(1405)$, $\eta(1475)$
or $\eta(1760)$. The positivity of the eigenvalues
allows only the last two candidates. For either
of these it is noted in section IV that there are
two solutions for the two quark vs four quark content
of the $\eta(958)$. The presumably favored solution
results in  $\eta(958)$ with a mainly two quark content,
while the less favored solution results in a mainly
four quark content for the $\eta(958)$.

    In sections V, VI and VII we gave a detailed proof that
the low energy theorem for pion pion scattering holds
in the present model with massless pions,
for any choice of chiral invariant potential.
The proof made use of the ``generating equations", stated
in section II, to relate the four particle, three particle
and two particle (ie mass term) vertices
to each other. We carried out this somewhat lengthy
calculation for two reasons.
First, since the pion in the model has
a non negligible,
though small four quark content, one might
wonder whether
the theorem actually does hold. Second, it
is expected to be useful
to calculate the scattering
amplitude, Eq.(\ref{wholeamp})
in the resonance region, rather than
close to threshold, as the theorem
requires.

    Clearly, there are a number of other
interesting directions
for further work. We plan to add mass terms
in the same systematic scheme employed
in section IV for selecting the most important
chiral invariant terms. Mixing with glueball
states and possibly other
chiral nonets is also an intriguing possibility.
Of course, an important ingredient to be taken into account
would be the changes in the model parameters
which result from  unitarizing the
tree level scattering amplitudes and comparing with the
unitarized amplitudes with experiment.
This was carried out for the 2 flavor Gell Mann-
Levy model in \cite{AS94} and for the 3 flavor single M model
in \cite{BFMNS01}.

\section*{Acknowledgments} \vskip -.5cm
We are happy to thank A. Abdel-Rehim, D. Black, M. Harada,
S. Moussa, S. Nasri and F. Sannino for many helpful
related discussions.
The work of A.H.F. has been partially supported by a 2006-2007
Crouse Grant from the School of Arts \& Sciences, SUNY
Institute of Technology.
The work of R.J. and J.S. is supported in part by the U. S. DOE under
Contract no. DE-FG-02-85ER 40231.

\appendix
\section{Some corrections}
    We have found the following minor corrections to ref \cite{FJS05}:

1. In Eq.(A1) the fifth term on the right hand side should properly read,
$d_2 {\rm Tr}(M'M'^\dagger)$.

2. In the sentence immediately following Eq.(A1), $d_2$ should be added
to the list of coefficients

which are $U(1)_A$ invariant.

3. In Eq.(19), the denominator of the argument of the ``ln" in the first
term should read ${\rm det} M^\dagger$.

4. In the last line of Eq.(58) the left hand side should read $\beta_3$.

5. In the last approximate equality in Eq.(60) the left hand side should
read $\beta_3$.

\section{Parameter determination}
Given the inputs: the pion decay constant, $F_\pi$; the mass of the
$a_0(980)$, $m_a$;
the mass of the  $a_0(1450)$, $m_{a'}$; the mass of the
$\pi(1300)$, $m_{\pi'}$, the independent model parameters
which don't involve the $U(1)_A$ violating terms can be
successively
determined (in the order given) by the equations:
\begin{eqnarray}
2 d_2 &=& {  {m_a^2 m_{a'}^2} \over {m_a^2 +m_{a'}^2 - m_{\pi'}^2} }
\nonumber \\
(\alpha e_3^a)^2 &=&\frac{1}{64} \left( (m_a^2 - m_{a'}^2)^2  - [4 d_2 -
(m_a^2
+m_{a'}^2)]^2\right)
\nonumber \\
4 c_2 &=&  m_a^2 + m_{a'}^2  - 2d_2 - {{ 56(\alpha e_3^a)^2} \over d_2}
\nonumber \\
{\beta\over \alpha} &=& {{-2 (\alpha e_3^a)} \over d_2}
\nonumber \\
\alpha^2 &=& {1\over 4}\,  {{F_\pi^2} \over {1+ (\beta/\alpha)^2}}
\nonumber \\
c_4^a &=& {1 \over {2 \alpha^2}}
\left( c_2 + {{8 (\alpha e_3^a)^2} \over d_2} \right)
\label{lagpara}
\end{eqnarray}

    The first equation tells us that $d_2$ is positive for
the experimental input masses. We take $\alpha$ and $\beta$ to
be positive. Then the fourth equation shows that $e_3^a$ must be negative.
Finally $c_2$ and $c_4^a$ will be positive.

    Once the above parameters are determined, the parameters $\gamma_1$
and $c_3$ of the $U(1)_A$ violating sector are obtained in terms of
the mass of the $\eta(958)$, $m_{\eta 1}$ and the mass of a suitable
heavier
$0^-$ isosinglet, $m_{\eta 2}$ as follows. First,
$\gamma_1$ is found as a solution of the quadratic equation:
\begin{eqnarray}
0&=&S\gamma_1^2+T\gamma_1+U,
\nonumber \\
S&=&r(4+\frac{\alpha^2}{\beta^2}),
\nonumber \\
T&=&r(4-2\frac{\alpha^2}{\beta^2}),
\nonumber \\
U&=&r(1+\frac{\alpha^2}{\beta^2})-36,
\nonumber \\
r&=&\frac{4m_{\eta
1}^2m_{\eta 2}^2}
  {y_0[m_{\eta 1}^2+ m_{\eta 2}^2-y_0(1+z_0^2)]}.
\label{findgamma1}
\end{eqnarray}
In addition,
\begin{equation}
c_3=-\frac{m_{\eta 1}^2m_{\eta 2}^2\alpha^2}{24y_0}
\label{findc3}
\end{equation}

Next we give the numerical values of the parameters for the
central values
of all the listed input masses
except for $m[\pi(1300)]$ which instead will take the typical
value allowed by both the data and
by the model, 1215 MeV. Table \ref{T_6param} shows the results
for the parameters
which are not associated with
the  $U(1)_A$ violating part of the Lagrangian.

\begin{table}[htbp]
\begin{center}
\begin{tabular}{c|c}
\hline \hline
$c_2 ({\rm GeV}^2)$    & 9.64 $\times 10^{-2}$ \\
$d_2  ({\rm GeV}^2)$   & 6.32 $\times 10^{-1}$\\
$e_3^a  ({\rm GeV})$   & $-2.14$\\
$c_4^a $               & 42.1  \\
$\alpha  ({\rm GeV})$  & 6.06 $\times 10^{-2}$\\
$\beta  ({\rm GeV})$   & 2.49 $\times 10^{-2}$
\\
\hline
\end{tabular}
\end{center}
\caption[]{
Calculated Lagrangian parameters:$c_2$, $d_2$, $e_3^a$, $c_4^a$
and vacuum values: $\alpha$, $\beta$.
}
\label{T_6param}
\end{table}

Table \ref{T_2param} shows the calculated Lagrangian
parameters associated with the $U(1)_A$ violating terms.
Two ``scenarios" associated with different identifications
of the heavy $\eta$ which is the partner of the $\eta(958)$
are shown (I assumes $\eta(1475)$ to be chosen while II
assumes $\eta(1760)$ to be chosen.) For each scenario, the
two solutions (labeled 1 and 2) are shown.

\begin{table}[htbp]
\begin{center}
\begin{tabular}{c|c|c|c|c}
\hline \hline
                   &       I1             & I2
                   &       II1            & II2\\
$c_3 ({\rm GeV}^4)$&$-2.42 \times 10^{-4}$&$-2.42 \times 10^{-4}$
                   &$-3.44 \times 10^{-4}$& $-3.44 \times 10^{-4}$\\
$\gamma_1$         & 5.38 $\times 10^{-1}$&  $2.53 \times 10^{-1}$
                   & 8.69 $\times 10^{-1}$& $-7.76 \times 10^{-2}$\\
\hline
\end{tabular}
\end{center}
\caption[]{
Calculated parameters: $c_3$ and $\gamma_1$.
}
\label{T_2param}
\end{table}

Using these parameters we next list the mixing matrices for,
respectively, the two $0^-$ octet states, the two $0^+$ octet
states and the two $0^+$ singlet states:

\begin{equation}
(R_\pi^{-1})  =
\left[
\begin{array}{cc}
0.925 & $0.380$ \\
-$0.380$ & 0.925\\
\end{array}
\right], \hspace{.3cm}
(L_a^{-1})  =
\left[
\begin{array}{cc}
-$0.496$ & $0.869$ \\
0.869    & 0.496\\
\end{array}
\right], \hspace{.3cm}
(L_0^{-1})  =
\left[
\begin{array}{cc}
0.711 & 0.703 \\
-$0.703$    & 0.711\\
\end{array}
\right].
\label{mms}
\end{equation}

Similarly, the mixing matrices for the two solutions for
scenario I of the $0^-$ singlet states are:

\begin{equation}
I\,\,1:(R_0^{-1})  =
\left[
\begin{array}{cc}
-$0.671$ & 0.742 \\
0.742 & 0.671\\
\end{array}
\right],     \hspace{.3cm}
I\,\,2: (R_0^{-1})=
\left[
\begin{array}{cc}
0.858 & -$0.514$ \\
0.514 & 0.858\\
\end{array}
\right].
\label{R0_num_I}
\end{equation}

Finally, the mixing matrices for the two solutions for
scenario II of the $0^-$ singlet states are:

\begin{equation}
II\,\,1:(R_0^{-1})  =
\left[
\begin{array}{cc}
-$0.413$ & 0.910 \\
0.910 & 0.413\\
\end{array}
\right], \hspace{.3cm}
II\,\,2: (R_0^{-1})  =
\left[
\begin{array}{cc}
0.974 & -$0.228$ \\
0.228 & 0.074\\
\end{array}
\right].
\label{R0_num_II}
\end{equation}

\end{document}